\documentclass[notitlepage,11pt]{article}
\usepackage{geometry}
\usepackage{mathrsfs}
\usepackage{lscape}
\usepackage{atlasphysics}
\usepackage{heppennames}
\usepackage{hepnicenames}
\usepackage{units}
\usepackage{url}
\usepackage{graphicx}
\DeclareGraphicsExtensions{.eps,.eps.gz,.pdf}
\usepackage{atlasphysics}
\usepackage{units}
\usepackage{xspace}
\usepackage{mathrsfs}
\usepackage{subfigure}
\usepackage{textpos}
\usepackage{arydshln}
\usepackage{multirow}
\usepackage{lineno}
\usepackage{placeins}
\usepackage{appendix}

\usepackage{jheppub}

\usepackage{preprintcover}
\PreprintCoverPaperTitle{Search for supersymmetry in events with
                      large missing transverse momentum, jets,
                      and at least one tau lepton in 20\,fb$\protect\boldmath{}^{-1}$
                      of $\sqrt{s}=$~8\,TeV proton--proton
                      collision data with the ATLAS detector}
\PreprintIdNumber{CERN-PH-EP-2014-144}  
\PreprintCoverAbstract{
A search for supersymmetry (SUSY) in events with large missing transverse
momentum, jets, at least one hadronically decaying tau lepton and zero or
one additional light leptons (electron/muon), has been performed using
20.3\,fb${}^{-1}$ of proton--proton collision data at $\sqrt{s} = 8$~TeV
recorded with the ATLAS detector at the Large Hadron Collider.
No excess above the Standard Model background expectation is observed
in the various signal regions and 95\% confidence level upper limits on
the visible cross section for new phenomena are set.
The results of the analysis are interpreted in several SUSY scenarios,
significantly extending previous limits obtained in the same final states.
In the framework of minimal gauge-mediated SUSY breaking models, values of
the SUSY breaking scale $\Lambda$ below 63~TeV are excluded,
independently of $\tan\beta$.
Exclusion limits are also derived for an mSUGRA/CMSSM model, in both the
$R$-parity-conserving and $R$-parity-violating case.
A further interpretation is presented in a framework of natural
gauge mediation, in which the gluino is assumed to be the only
light coloured sparticle and gluino masses below 1090~GeV are excluded.
}
\PreprintJournalName{JHEP}
\newcommand{\Alpgen}{{\tt ALPGEN}\xspace}
\newcommand{\Sherpa}{{\tt SHERPA}\xspace}
\newcommand{\Jimmy}{{\tt JIMMY}\xspace}
\newcommand{\Mcatnlo}{{\tt MC@NLO}\xspace}
\newcommand{\Powheg}{{\tt POWHEG}\xspace}

\newcommand{\Herwig}{{\tt HERWIG}\xspace}
\newcommand{\Herwigpp}{{\tt Herwig++}\xspace}

\newcommand{\Tauola}{{\tt TAUOLA}\xspace}

\newcommand{\Pythia}{{\tt PYTHIA}\xspace}

\newcommand{\Acermc}{{\tt AcerMC}\xspace}

\renewcommand{\ttbar}{\antibar{t}\xspace}
\newcommand{\neutralino}{\ensuremath{\tilde{\chi}^{0}_{1}}\xspace}
\newcommand{\neutralinotwo}{\ensuremath{\tilde{\chi}^{0}_{2}}\xspace}
\newcommand{\neutralinoi}{\ensuremath{\tilde{\chi}^{0}_{i}}\xspace}
\newcommand{\chargino}{\ensuremath{\tilde{\chi}^{\pm}_{1}}\xspace}

\newcommand{\gravitino}{\ensuremath{\tilde{G}}\xspace}
\newcommand{\sneutrino}{\ensuremath{\tilde{\nu}}\xspace}

\newcommand{\Nfive}{\ensuremath{N_{5}}\xspace}
\newcommand{\Mmess}{\ensuremath{M_{\mathrm{mess}}}\xspace}
\newcommand{\Cgrav}{\ensuremath{C_{\mathrm{grav}}}\xspace}

\newcommand{\CL}{CL\xspace}
\renewcommand{\met}{\ensuremath{E_{\mathrm{T}}^{\mathrm{miss}}}\xspace}
\newcommand{\ptmiss}{\ensuremath{{p}_\mathrm{T}^\mathrm{miss}}\xspace}
\newcommand{\ptmissv}{\ensuremath{\vec{{p}}_\mathrm{T}^\mathrm{~miss}}\xspace}
\newcommand{\meff}{\ensuremath{m_{\mathrm{eff}}}\xspace}
\renewcommand{\pt}{\ensuremath{p_{\mathrm{T}}}\xspace}

\renewcommand{\mass}[1]{\ensuremath{m_{#1}}}
\newcommand{\mT}{\ensuremath{\mass{\mathrm T}^{\tau}}\xspace}
\newcommand{\mTone}{\ensuremath{\mass{\mathrm T}^{\tau_1}}\xspace}

\newcommand{\mTT}{\ensuremath{\mass{\mathrm T}^{\tau_1}+\mass{\mathrm T}^{\tau_2}}\xspace}
\newcommand{\mTTkerned}{\ensuremath{\mass{\mathrm T}^{\tau_1}\!+\!\mass{\mathrm T}^{\tau_2}}\xspace}
\newcommand{\HTtj}{\ensuremath{H_{\mathrm{T}}^\text{2j}}\xspace}

\newcommand{\sect}{section}
\newcommand{\Fig}{Figure\xspace} 
\newcommand{\Figs}{Figures\xspace} 
\newcommand{\fig}{figure\xspace}

\newcommand{\refer}{ref.\xspace}

\newcommand{\Tab}{Table\xspace} 
\newcommand{\Tabs}{Tables\xspace} 
\newcommand{\tab}{table\xspace}
\newcommand{\tabs}{tables\xspace}

\newcommand{\onetau}{\mbox{$1\tau$}\xspace}
\newcommand{\twotau}{\mbox{$2\tau$}\xspace}
\newcommand{\taumu}{\mbox{$\tau\!+\!\mu$}\xspace}
\newcommand{\tauel}{\mbox{$\tau\!+\!e$}\xspace}
\newcommand{\taulep}{\mbox{$\tau\!+\!{}$lepton}\xspace}
\newcommand{\tauleps}{\mbox{$\tau\!+\!\ell$}\xspace}
\newcommand{\LJetPt}{\ensuremath{\pt^\mathrm{jet1}}\xspace}
\newcommand{\SLJetPt}{\ensuremath{\pt^\mathrm{jet2}}\xspace}

\newcommand{\njet}{\ensuremath{N_\text{jet}}\xspace}
\newcommand{\moh}{\ensuremath{m_{{1}\kern-0.08em/\kern-0.07em{2}}}\xspace}
\newcommand{\mz}{\ensuremath{m_0}\xspace}
\newcommand{\nbjet}{\ensuremath{N_{b\text{-jet}}}\xspace}

\renewcommand{\GeV}[1]{\ensuremath{\unit[#1]{GeV}} }
\renewcommand{\TeV}[1]{\ensuremath{\unit[#1]{TeV}} }

\newcommand{\mtlep}{\ensuremath{m^\ell_{\mathrm{T}}}\xspace}

\newcommand{\mgl}{\ensuremath{m_{\gluino}}\xspace}
\newcommand{\mst}{\ensuremath{m_{\stau}}\xspace}

\newcommand{\integLumi}{\unit[20.3]{\ifb}\xspace}
\newcommand{\integLumiE}{\unit[$(20.3\pm 0.6)$]{\ifb}\xspace}
\newcommand{\onetauLooseQcdEstim}{$1.12\pm0.49\,{}^{+1.27}_{-1.12}$}
\newcommand{\onetauLooseWEstim}{$3.13\pm0.57\pm1.10$}
\newcommand{\onetauLooseZEstim}{$1.89\pm0.56\pm1.58$}
\newcommand{\onetauLooseTopEstim}{$3.87\pm0.99\pm1.62$}
\newcommand{\onetauLooseDBEstim}{$0.47\pm0.18\pm0.16$}
\newcommand{\onetauLooseTotalEstim}{$10.5\pm1.4\pm2.6$}
\newcommand{\onetauLooseData}{$12$}

\newcommand{\onetauTightQcdEstim}{$0.23\pm0.10\pm0.24$}
\newcommand{\onetauTightWEstim}{$0.73\pm0.20\pm0.69$}
\newcommand{\onetauTightZEstim}{$0.42\pm0.15\pm0.14$}
\newcommand{\onetauTightTopEstim}{$0.82\pm0.34\pm0.46$}
\newcommand{\onetauTightDBEstim}{$0.16\pm0.10\pm0.09$}
\newcommand{\onetauTightTotalEstim}{$2.4\pm0.4\pm0.8$} 
\newcommand{\onetauTightData}{$3$}
\newcommand{\onetauTightGMSB}{$6.4\pm 0.7\pm 0.4$} 
\newcommand{\onetauTightmSUGRA}{$15.7\pm 2.2\pm 1.1$}
\newcommand{\onetauTightbRPV}{$2.8\pm 0.4\pm 0.4$} 

\newcommand{\ditauQCDInclusive}{\ensuremath{0.12\pm0.05\pm0.06}}
\newcommand{\ditauQCDGMSB}{\ensuremath{0.062\pm0.045\pm0.021}}
\newcommand{\ditauQCDnGM}{\ensuremath{0.066\pm0.045\pm0.032}}
\newcommand{\ditauQCDbRPVmoh}{\ensuremath{0.11\pm0.05\pm0.04}}

\newcommand{\ditauInclusiveScaleStatSystNoLabel}{\ensuremath{2.9\pm0.4\pm0.7}} 

\newcommand{\ditauInclusiveData}{\ensuremath{3}}

\newcommand{\ditauGMSBScaleStatSystNoLabel}{\ensuremath{0.28\pm0.10\pm0.22}}

\newcommand{\ditauGMSBData}{\ensuremath{0}}

\newcommand{\ditaunGMScaleStatSystNoLabel}{\ensuremath{3.1\pm0.5\pm0.9}} 

\newcommand{\ditaunGMData}{\ensuremath{1}}

\newcommand{\ditaubRPVmohScaleStatSystNoLabel}{\ensuremath{1.09\pm0.19\pm0.39}}

\newcommand{\ditaubRPVmohData}{\ensuremath{1}}

\newcommand{\ditauInclusiveDB}{\ensuremath{0.39\pm0.19\pm0.30}}
\newcommand{\ditauInclusiveTop}{\ensuremath{0.57\pm0.14\pm0.32}}
\newcommand{\ditauInclusiveW}{\ensuremath{1.26\pm0.33\pm0.54}}
\newcommand{\ditauInclusiveZ}{\ensuremath{0.54\pm0.15\pm0.64}}

\newcommand{\ditauGMSBDB}{\ensuremath{0\pm0}}
\newcommand{\ditauGMSBTop}{\ensuremath{0.050\pm0.031\pm0.053}}
\newcommand{\ditauGMSBW}{\ensuremath{0.14\pm0.07\pm0.18}}
\newcommand{\ditauGMSBZ}{\ensuremath{0.037\pm0.020\pm0.042}}

\newcommand{\ditaunGMDB}{\ensuremath{0\pm0}}
\newcommand{\ditaunGMTop}{\ensuremath{1.65\pm0.38\pm0.65}}
\newcommand{\ditaunGMW}{\ensuremath{0.78\pm0.31\pm0.47}}
\newcommand{\ditaunGMZ}{\ensuremath{0.65\pm0.28\pm0.94}}

\newcommand{\ditaubRPVmohDB}{\ensuremath{0\pm0}}
\newcommand{\ditaubRPVmohTop}{\ensuremath{0.32\pm0.10\pm0.19}}
\newcommand{\ditaubRPVmohW}{\ensuremath{0.48\pm0.15\pm0.31}}
\newcommand{\ditaubRPVmohZ}{\ensuremath{0.18\pm0.07\pm0.21}}

\newcommand{\GMSBLimitAll}{\unit[63]{TeV}\xspace}
\newcommand{\GMSBLimitBestTanBeta}{\unit[73]{TeV}\xspace}
\newcommand{\GMSBLimitGluino}{\unit[1600]{GeV}\xspace}
\newcommand{\nGMLimitAll}{\unit[1090]{GeV}\xspace}
\newcommand{\bRPVmohlimit}{\unit[680]{GeV}\xspace}
\newcommand{\bRPVmzlimit}{\unit[920]{GeV}\xspace}
\newcommand{\bRPVmzlimitposition}{\unit[360]{GeV}\xspace}
\newcommand{\mSugramohlimitsmall}{\unit[640]{GeV}\xspace}
\newcommand{\mSugramohlimitlarge}{\unit[300]{GeV}\xspace}

\newcommand{\taumuGMSBData}{\ensuremath{2}}
\newcommand{\taumuNGMData}{\ensuremath{2}}
\newcommand{\taumuBRPVData}{\ensuremath{7}}
\newcommand{\taumuMSugData}{\ensuremath{9}}

\newcommand{\tauelGMSBData}{\ensuremath{1}}
\newcommand{\tauelNGMData}{\ensuremath{8}}
\newcommand{\tauelBRPVData}{\ensuremath{3}}
\newcommand{\tauelMSugData}{\ensuremath{14}}

\newcommand{\tauelGMSBAll}{\ensuremath{1.34\pm0.33 \pm 0.80}}
\newcommand{\tauelGMSBDiBosons}{\ensuremath{0.29\pm0.13 \pm 0.28}}

\newcommand{\tauelGMSBTop}{\ensuremath{0.52\pm0.26 \pm 0.54}}
\newcommand{\tauelGMSBW}{\ensuremath{0.25\pm0.11 \pm 0.31}}
\newcommand{\tauelGMSBZ}{\ensuremath{0.28\pm0.12 \pm 0.29}}

\newcommand{\tauelGMSBQCD}{\ensuremath{<0.2}}

\newcommand{\taumuGMSBAll}{\ensuremath{0.98\pm0.31 \pm 0.35}}

\newcommand{\taumuGMSBTop}{\ensuremath{0.02\pm0.02 \pm 0.01}}
\newcommand{\taumuGMSBW}{\ensuremath{0.32\pm0.13 \pm 0.08}}
\newcommand{\taumuGMSBZ}{\ensuremath{0.33\pm0.24 \pm 0.5}}
\newcommand{\taumuGMSBDiBoson}{\ensuremath{0.29\pm0.13 \pm 0.16}}

\newcommand{\taumuGMSBQCD}{\ensuremath{<0.01} }

\newcommand{\tauelNGMAll}{\ensuremath{4.3\pm0.9\pm2.0}} 

\newcommand{\tauelNGMTop}{\ensuremath{2.98 \pm 0.82 \pm 1.93}}
\newcommand{\tauelNGMW}{\ensuremath{0.45 \pm 0.14 \pm 0.28}}
\newcommand{\tauelNGMZ}{\ensuremath{0.11 \pm 0.06 \pm 0.11}}
\newcommand{\tauelNGMQCD}{\ensuremath{<0.1}}
\newcommand{\tauelNGMDiBoson}{\ensuremath{ 0.73 \pm 0.21 \pm 0.21}}

\newcommand{\taumuNGMAll}{\ensuremath{3.6\pm0.9\pm1.2}} 

\newcommand{\taumuNGMTop}{\ensuremath{2.80 \pm 0.83\pm 0.97}}
\newcommand{\taumuNGMW}{\ensuremath{0.39 \pm 0.15\pm 0.42}}
\newcommand{\taumuNGMZ}{\ensuremath{0.06 \pm 0.03\pm 0.07}}
\newcommand{\taumuNGMQCD}{\ensuremath{<0.04}}
\newcommand{\taumuNGMDiBoson}{\ensuremath{0.32 \pm 0.14\pm 0.28}}

\newcommand{\tauelBRPVAll}{\ensuremath{4.0\pm0.8\pm1.3}} 
\newcommand{\tauelBRPVDiBosons}{\ensuremath{0.22\pm0.12\pm 0.12}}

\newcommand{\tauelBRPVTop}{\ensuremath{1.99\pm0.59\pm 0.81}}
\newcommand{\tauelBRPVW}{\ensuremath{1.61\pm0.54\pm0.58}}
\newcommand{\tauelBRPVZ}{\ensuremath{0.20\pm0.09\pm0.81}}
\newcommand{\tauelBRPVQCD}{\ensuremath{<0.3}}

\newcommand{\taumuBRPVAll}{\ensuremath{2.5\pm0.6\pm1.0}} 
\newcommand{\taumuBRPVDiBoson}{\ensuremath{0.22\pm0.12\pm 0.10}}

\newcommand{\taumuBRPVTop}{\ensuremath{1.22\pm0.46\pm 0.57}}
\newcommand{\taumuBRPVW}{\ensuremath{0.82\pm0.32\pm 0.70}}
\newcommand{\taumuBRPVZ}{\ensuremath{ 0.29\pm0.13\pm 0.16}}
\newcommand{\taumuBRPVQCD}{\ensuremath{<0.04}}

\newcommand{\tauelMSugAll}{\ensuremath{10.0\pm1.4\pm3.0}} 
\newcommand{\tauelMSugDiBosons}{\ensuremath{1.47\pm0.30\pm 0.32}}

\newcommand{\tauelMSugTop}{\ensuremath{7.43\pm1.31\pm 2.52}}
\newcommand{\tauelMSugW}{\ensuremath{0.96\pm0.22\pm 0.46}}
\newcommand{\tauelMSugZ}{\ensuremath{0.15\pm0.07\pm 0.16}}
\newcommand{\tauelMSugQCD}{\ensuremath{<0.4}}

\newcommand{\taumuMSugAll}{\ensuremath{9.9\pm1.5\pm3.3}} 
\newcommand{\taumuMSugDiBoson}{\ensuremath{0.72\pm0.21\pm 0.55}}

\newcommand{\taumuMSugTop}{\ensuremath{8.36\pm1.40\pm 2.90}}

\newcommand{\taumuMSugW}{\ensuremath{0.75\pm0.20\pm 0.38}}
\newcommand{\taumuMSugZ}{\ensuremath{0.07\pm0.03\pm 0.07}}
\newcommand{\taumuMSugQCD}{\ensuremath{<0.01}}

\newcommand{\GMSBBenchPars}{$\Lambda\!=\!\TeV{60}\;/\;\tan\beta\!=\!30$\xspace}
\newcommand{\nGMBenchPars}{$\mgl\!=\!\GeV{940}\;/\;\mst\!=\!\GeV{210}$\xspace}
\newcommand{\bRPVBenchPars}{$\mz\!=\!\GeV{600}\;/\;\moh\!=\!\GeV{600}$\xspace}
\newcommand{\mSugraBenchPars}{$\mz\!=\!\GeV{800}\;/\;\moh\!=\!\GeV{400}$\xspace}

\newcommand{\twotauGMSBBench}{$9.7\pm0.8\pm0.6$} 
\newcommand{\twotaunGMBench}{$17.7\pm0.8\pm1.1$} 
\newcommand{\twotaubRPVBench}{$1.9\pm0.3\pm0.2$} 

\newcommand{\taumuGMSBBench}{$4.34\pm0.48\pm0.26$}
\newcommand{\taumuNGMBench}{$5.2\pm0.4\pm0.4$} 
\newcommand{\taumuMSUGRABench}{$13.6\pm2.0\pm0.5$} 
\newcommand{\taumuBRPVBench}{$5.55\pm0.52\pm0.24$}

\newcommand{\tauelGMSBBench}{$8.1\pm0.5\pm1.0$} 
\newcommand{\tauelNGMBench}{$6.4\pm0.5\pm0.5$} 
\newcommand{\tauelMSUGRABench}{$14.1\pm1.9\pm0.8$} 
\newcommand{\tauelBRPVBench}{$4.03\pm0.48\pm0.18$}

\abstract{
A search for supersymmetry (SUSY) in events with large missing transverse
momentum, jets, at least one hadronically decaying tau lepton and zero or
one additional light leptons (electron/muon), has been performed using
20.3\,fb${}^{-1}$ of proton--proton collision data at $\sqrt{s} = 8$~TeV
recorded with the ATLAS detector at the Large Hadron Collider.
No excess above the Standard Model background expectation is observed
in the various signal regions and 95\% confidence level upper limits on
the visible cross section for new phenomena are set.
The results of the analysis are interpreted in several SUSY scenarios,
significantly extending previous limits obtained in the same final states.
In the framework of minimal gauge-mediated SUSY breaking models, values of
the SUSY breaking scale $\Lambda$ below 63~TeV are excluded,
independently of $\tan\beta$.
Exclusion limits are also derived for an mSUGRA/CMSSM model, in both the
$R$-parity-conserving and $R$-parity-violating case.
A further interpretation is presented in a framework of natural
gauge mediation, in which the gluino is assumed to be the only
light coloured sparticle and gluino masses below 1090~GeV are excluded.
}
\title{Search for supersymmetry in events with large missing transverse momentum,
       jets, and at least one tau lepton in 20\,fb${}^{-1}$
       of $\sqrt{s}=$~8\,TeV proton--proton
       collision data with the ATLAS detector}
\author{The ATLAS collaboration}
\emailAdd{atlas.publications@cern.ch} 
\begin{document}

\maketitle

\section{Introduction}
\label{sec:intro}
Supersymmetry (SUSY)~\cite{Golfand:1971iw,Neveu:1971rx,Ramond:1971gb,Volkov:1973ix,Wess:1974tw} 
introduces a symmetry between fermions and bosons, resulting in a SUSY
partner (sparticle) for each Standard Model (SM) particle, with
identical mass and quantum numbers except a difference of half a unit
of spin. As none of these sparticles have been observed with the
same mass as their SM partners, SUSY must be
a broken symmetry if realized in nature.  Assuming $R$-parity
conservation~\cite{Fayet:1976et,Fayet:1977yc,Farrar:1978xj,Fayet:1979sa,Dimopoulos:1981zb},
sparticles are produced in pairs and then decay through cascades involving
other sparticles until the lightest SUSY particle (LSP), which is stable, is produced.
In many SUSY models tau leptons can provide an
important signature for new physics.
Naturalness arguments~\cite{Barbieri:1987fn,deCarlos:1993yy} suggest that the lightest third-generation sparticles should have 
masses of a few hundred GeV to protect the Higgs boson mass from quadratically divergent 
quantum corrections. Light sleptons could play a role in the co-annihilation of 
neutralinos in the early universe, and, in particular, models with light tau sleptons (staus) are consistent
with dark matter searches \cite{Vasquez:2011}.
If squarks and gluinos, superpartners of quarks and gluons,%
\footnote{In addition to squarks and gluinos, charged sleptons
and sneutrinos are superpartners of charged leptons and neutrinos.
The SUSY partners of the gauge and Higgs bosons are called gauginos and higgsinos, respectively.
The charged, electroweak gauginos and higgsinos mix to form charginos
($\chi_{i}^{\pm}$, $i$ = 1,2), and the neutral ones mix to form neutralinos
($\chi_{j}^{0}$, $j$ = 1,2,3,4).
Finally the gravitino is the SUSY partner of the graviton.
} are in the LHC reach, their production rate may be dominant among SUSY processes. They could then decay in cascades 
involving tau leptons, high transverse momentum jets and missing transverse momentum from the LSP, which escapes undetected.
More details about the various SUSY models considered in this paper are given in \sect~\ref{sec:susy}.
Furthermore, should SUSY or any other theory of physics Beyond the Standard Model (BSM) be discovered, independent 
studies of all three lepton flavours are necessary to investigate the coupling structure of the new physics, 
especially with regard to lepton universality. 

This paper reports on an inclusive search for SUSY particles produced via the strong interaction
in events with large missing transverse momentum, jets and at least one
hadronically decaying tau lepton.
Four distinct topologies are studied: one tau lepton (``\onetau'')
or two or more tau leptons (``\twotau'') in the final state, with no
additional light leptons (e/$\mu$); 
or one or more tau leptons with exactly one electron (``\tauel'') or
muon (``\taumu''). These orthogonal channels have been optimized
separately, and, where relevant, are statistically combined to increase the analysis sensitivity.
The analysis is performed using \integLumi of proton--proton ($pp$)
collision data at $\sqrt{s}=\unit[8]{TeV}$ recorded with the ATLAS
detector at the Large Hadron Collider (LHC) in the 2012 run.  The results are 
interpreted in several different models, which are described in more detail in \sect~\ref{sec:susy}:
a minimal gauge-mediated supersymmetry breaking
(GMSB) model~\cite{Dine:1981gu,AlvarezGaume:1981wy,Nappi:1982hm,Dine:1993yw,Dine:1994vc,Dine:1995ag}, 
an mSUGRA/CMSSM~\cite{Chamseddine:1982,Barbieri:1982eh,Ibanez:1982ee,Hall:1983iz,Ohta:1982wn,Kane:1993td} model, 
a natural gauge mediation framework (nGM)~\cite{Barnard:2012au} and a bilinear $R$-parity-violation 
(bRPV)~\cite{Hirsch:2000bn,deCampos:2007bn} model.

Previous searches for direct production of the SUSY partners of the tau lepton in the minimal GMSB model have been reported
by the LEP Collaborations ALEPH~\cite{Heister:2002vh},
DELPHI~\cite{Abdallah:2002rd} and OPAL~\cite{Abbiendi:2005gc}.
The analysis reported in this paper extends the searches presented in \refer~\cite{ATLAS:SusyTau5fb}.
The CMS Collaboration presented the results of a supersymmetry search in events with tau leptons, jets and missing transverse momentum in 
4.98 fb$^{-1}$ of 7 TeV data in \refer~\cite{Chatrchyan:2013dsa}. 

\section{SUSY scenarios}
\label{sec:susy}
The search presented in this paper is sensitive to a variety of SUSY scenarios,
which are outlined below.
In particular, good sensitivity is achieved for SUSY strong production processes due to the requirement of 
several high-momentum jets. 
\vspace{5pt}

{\bf GMSB model -} Minimal GMSB models can be described by six parameters: the SUSY-breaking mass
scale in the low-energy sector ($\Lambda$), the messenger mass
(\Mmess), the number of SU(5) messenger fields (\Nfive), the ratio of the
vacuum expectation values of the two Higgs doublets ($\tan\beta$), the
Higgs sector mass parameter ($\mu$) and the scale factor for the
gravitino mass (\Cgrav). For the analysis presented here,
$\Lambda$ and $\tan\beta$ are treated as free parameters, 
and the other parameters are fixed to
the values used in \refer~\cite{ATLAS:SusyTau5fb}:
$\Mmess=\unit[250]{TeV}$, $\Nfive=3$, $\mu>0$ and $\Cgrav=1$. 
With this choice of parameters, the production of squark and/or gluino pairs is
expected to dominate over other SUSY processes at the LHC.

These sparticles decay into the next-to-lightest SUSY particle (NLSP),
which subsequently decays to the LSP.
In gauge-mediated models, the LSP is always a very light gravitino (\gravitino). 
The experimental signatures are largely determined by the nature of
the NLSP: this can be either the lightest stau (\stau), a 
selectron or a smuon (\slepton), the lightest neutralino (\neutralino), or
a sneutrino (\sneutrino), leading to final states usually containing tau leptons,
light leptons ($\ell=e,\mu$), photons, or neutrinos, respectively.
In most of the GMSB parameter space considered here the \stau ~is
the NLSP for large values of $\tan\beta$ ($\tan\beta >$ 20),
and final states contain  between two and four tau leptons.
In the region where the mass difference between the \stau ~and
the \slepton ~is smaller  than the sum of the tau and the light
lepton masses both the \stau ~and the \slepton ~decay directly
into the LSP and therefore both define the phenomenology.
\vspace{5pt}

{\bf mSUGRA/CMSSM model -} The mSUGRA/CMSSM scenario is defined by five parameters:
the universal scalar mass (\mz),
the universal trilinear coupling ($A_0$)
the universal gaugino mass (\moh), 
$\tan\beta$ and $\mu$.
These are chosen such that across a large area of the (\mz, \moh) plane
the mSUGRA/CMSSM lightest Higgs boson mass is compatible with the observed
mass of the recently discovered Higgs boson at the LHC \cite{ATLAS:Higgs,CMS:Higgs}.
Near the low \mz boundary of this area
the difference in mass between the \stau ~and the lightest SUSY particle,
the neutralino, is small and allows the two particles to co-annihilate in
the early universe \cite{Ellis:1999mm}. The dark matter relic density is
therefore brought down to values compatible with the Planck and WMAP
measurements~\cite{Planck:2013,Hinshaw:2012}.
The consequence of the small difference in
mass for the experimental sensitivity is a bias towards very low
momenta of at least one tau lepton and consequently towards fewer
detectable tau candidates in the final state.
\vspace{5pt}

{\bf nGM model -} A rich phenomenology is obtained in the framework of general gauge mediation (GGM)~\cite{Buican:2008ws}. 
Starting from GGM, it is possible to 
construct a set of natural Gauge Mediated (nGM) models where the phenomenology depends on the nature of the NLSP~\cite{Barnard:2012au, Asano:2010ma}.
Various models assume that the fermion mass hierarchies are generated by the same physics responsible for breaking SUSY (see for example~\cite{Gabella:2007mg} and~\cite{Craig:2012ng}).  
Typically in these models the entire third generation of sfermions is
lighter than the other two. Coupled with the fact that sleptons only get
soft masses through hypercharge interactions in gauge mediation, this 
leads to a stau NLSP. 
In the model considered here it is also assumed that the gluino is the only light coloured sparticle.
All squark and slepton mass parameters are set to \unit[2.5]{TeV} except the lightest stau mass, $m_{\tilde{\tau}}$, which is assumed to be smaller to allow 
a stau NLSP (this has no effect on the fine tuning). The bino and wino masses ($M_1$ and $M_2$ respectively) are also set to \unit[2.5]{TeV} while all trilinear 
coupling terms are set to zero. It is further assumed that $\mu \ll M_1,M_2$. 
This leaves the gluino mass $M_3$ and the stau mass $m_{\tilde{\tau}}$ as the only free parameters, if $\mu$ is also fixed.
The value of $\mu$ is set to \unit[400]{GeV} to ensure that strong production is the dominant process at the LHC; moreover, this choice of the $\mu$ parameter 
drives the mass of the $\chargino$, $\neutralino$ and $\neutralinotwo$, which are almost mass degenerate. 

The only light sparticles in the model are the stau, a light gluino, higgsino-dominated charginos and neutralinos, and a light gravitino, which is the LSP. 
Several decay modes are possible for the gluino:

\begin{enumerate}
\item $\tilde{g} \rightarrow g $\neutralinoi $\rightarrow g \tau$\stau $\rightarrow g \tau \tau \tilde{G}$, with $i=1,2$
\item $\tilde{g} \rightarrow q \bar{q} $\neutralinoi $\rightarrow q \bar{q} \tau$\stau $\rightarrow q \bar{q} \tau \tau \tilde{G}$, with $i=1,2$
\item $\tilde{g} \rightarrow q q' $\chargino $\rightarrow q q' \nu_{\tau}$\stau $\rightarrow q q' \nu_{\tau} \tau \tilde{G}$
\end{enumerate}

\noindent
where $q$ and $\bar{q}$ are almost exclusively quarks of heavy flavour (either top or bottom quarks). 
The first process proceeds through a squark-quark loop, and equal amounts of \neutralino and 
\neutralinotwo production are expected. The second and third processes proceed via an off-shell squark, 
and the relative proportion of the first process to the other two depends on the 
precise relationship between $M_3$ and the squark masses. At the lowest values of $M_3$, the first 
process dominates entirely. The effect of the last two processes increases with rising gluino mass 
(with $M_3$ approaching the squark masses). For $M_3$\,{\small$\gtrsim$}\,\unit[1]{TeV}, the proportion of decays through 
the first process is at the level of a few percent, and the other two processes are expected to dominate~\cite{Barnard:2012au}.
The branching ratios are approximately constant as a function of $M_3$ for the signal scenarios considered.

In gauge-mediated SUSY scenarios
a variety of mechanisms exist~\cite{Craig:2011yk, Auzzi:2011eu, Csaki:2012fh, Larsen:2012rq, Craig:2012hc} 
to generate a Higgs boson mass compatible with the observed value~\cite{ATLAS:Higgs,CMS:Higgs},
without changing the phenomenology of the models considered in this search. 
In the model used in this analysis, the lightest Higgs boson mass is specifically set to \unit[125]{GeV.}
\vspace{5pt}

{\bf bRPV model -} In the bRPV scenario, bilinear $R$-parity-violating (RPV) terms are assumed to be present in the superpotential, resulting in an unstable LSP. 
The RPV couplings are included in the mSUGRA/CMSSM model described above and, for a chosen
set of mSUGRA/CMSSM parameters, the bilinear RPV parameters are determined under the tree-level dominance scenario~\cite{Grossman:2004bn} by fitting 
them to neutrino oscillation data as described in \refer~\cite{Carvalho:2002bn}.
The neutralino LSP decays promptly through decay modes
that include neutrinos~\cite{Hirsch:2001bn}. The main LSP decay modes considered are:
\begin{enumerate}
\item $\neutralino \rightarrow W^{(*)} \mu$ (or $\tau$),
\item $\neutralino \rightarrow Z^{(*)}/h^{(*)} \nu$.
\end{enumerate}
\noindent
These result in final states with several leptons and jets, but a reduced missing transverse momentum compared with the standard $R$-parity-conserving
mSUGRA/CMSSM model.

\section{The ATLAS detector and data sample}
\label{sec:atlas_detector}
The ATLAS experiment is described in detail in \refer~\cite{Aad:2008zzm}.
It is a multi-purpose detector with a forward-backward symmetric cylindrical
geometry and nearly 4$\pi$ solid angle coverage.\footnote{ATLAS uses a right-handed coordinate
  system with its origin at the nominal interaction point (IP) in the
  centre of the detector and the $z$-axis along the beam pipe. The
  $x$-axis points from the IP to the centre of the LHC ring and the
  $y$-axis points upward. Cylindrical coordinates $(r,\phi)$ are used
  in the transverse plane, $\phi$ being the azimuthal angle around the
  beam pipe. The pseudorapidity is defined in terms of the polar angle
  $\theta$ as $\eta=-\ln\tan(\theta/2)$.}
The inner tracking detector (ID), covering $|\eta|<2.5$, consists
of a silicon pixel detector, a semiconductor microstrip detector and a
transition radiation tracker. The ID is surrounded by a thin
superconducting solenoid providing an axial \unit[2]{T} magnetic field and by
a fine-granularity lead/liquid-argon (LAr) electromagnetic calorimeter (covering $|\eta|<3.2$).
An iron/scintillator-tile calorimeter provides hadronic coverage in
the central pseudorapidity range ($|\eta|<1.7$). The endcap and
forward regions ($1.5<|\eta|<4.9$) are instrumented with LAr
calorimeters, with either steel, copper or tungsten as the absorber material, for
both the electromagnetic and hadronic measurements.
An extensive muon spectrometer system that incorporates large
superconducting toroidal air-core magnets surrounds the calorimeters. Three layers of precision gas chambers 
provide tracking coverage in the range $|\eta|<2.7$, while dedicated fast chambers allow triggering in the region $|\eta|<2.4$.

The data used in this search are $pp$ collisions recorded by the ATLAS detector at
a centre-of-mass energy of $\sqrt{s}=\unit[8]{TeV}$ during the
period from April 2012 to December 2012.
After the application of beam, detector and data-quality requirements,
the total integrated luminosity amounts to \integLumiE{}.
The luminosity measurement
is performed using techniques similar to those in \refer~\cite{Aad:2013ucp}, and
the calibration of the luminosity scale is derived from
beam-separation scans performed in November 2012. 
In the \onetau and \twotau channels, candidate events are triggered by
requiring a jet with high transverse momentum (\pt) and high 
missing transverse momentum (whose magnitude is denoted by \met)~\cite{Aad:2011xs}.
In the \tauel channel, candidate events are triggered
by requiring the presence of an energy cluster in the
electromagnetic calorimeter with a shower shape consistent with
that of an electron, and with uncorrected transverse energy ($\ET$) above \unit[24]{GeV}.
The selection is further refined by matching the cluster to an isolated track in
the ID~\cite{Aad:2011xs}. 
In order to maximize the efficiency for high-\pt electrons, 
data selected using a single-electron trigger with $\ET>\unit[60]{GeV}$ but no
isolation requirements are added.
In the \taumu channel, events are selected by requiring a muon candidate
identified as a single isolated track reconstructed by the ID and the muon spectrometer, 
with uncorrected transverse momentum above \unit[24]{GeV}. In addition, events are also selected
using a non-isolated muon trigger, with a muon \pt threshold of \unit[36]{GeV}~\cite{Aad:2011xs}.
The trigger requirements have been optimized to ensure a uniform trigger efficiency for
all data-taking periods, which exceeds 98\% with respect to the offline selection for 
all final states considered.

\section{Simulated samples}
\label{sec:SimulatedSamples}
%
Samples of Monte Carlo (MC) simulated events are used for evaluating the expected
SM backgrounds and for estimating the signal efficiencies for the different SUSY
models.
%
Samples of $W$+jets and $Z$+jets events with up to four jets from matrix elements (ME)
are simulated by the \Sherpa~\cite{Sherpa} generator version 1.4.1, where the
\texttt{CT10}~\cite{Lai:2010} set of parton distribution functions (PDFs) 
is used. To improve the agreement between
data and simulation, $W$/$Z$+jets events are reweighted based on the \pt of the
vector boson using measured $Z$ boson \pt distributions in the data \cite{ATLAS:2014stopcharm}.
For the purpose of evaluating generator uncertainties, additional $W$/$Z$+jets samples are
produced with the \Alpgen~2.14~\cite{Mangano:2002ea} MC generator,
which simulates $W$ and $Z/\gamma^*$ production with up to five accompanying
partons using the \texttt{CTEQ6L1} \cite{Pumplin:2002vw} set of PDFs.
$Z/\gamma^*$ events with $m_{\ell \ell}<$ \unit[40]{GeV} are referred to in this paper as ``Drell--Yan".
In the \Alpgen samples fragmentation and hadronization are performed with \Herwig~6.520~\cite{Corcella:2000bw}, using \Jimmy~\cite{Butterworth:1996zw} for the underlying
event simulation. The \Sherpa MC generator is used for simulating the production of
diboson events ($WW$, $WZ$ and $ZZ$). Alternative samples for evaluating
systematic uncertainties are generated by 
\Powheg~r2129~\cite{Nason:2004,Frixione:2007,Aioli:2010}
interfaced to \Pythia~8.165~\cite{Sjostrand:2007}. 

Top quark pair production is simulated with \Powheg~r2129
interfaced to \Pythia~6.426~\cite{Sjostrand:2006},
using the \texttt{CT10} PDF set.
To improve the agreement between data and simulation, \ttbar events
are reweighted based on the \pt of the \ttbar system; the weights
are extracted from the ATLAS measurement of the \ttbar differential 
cross section at $\sqrt{s}=\unit[7]{TeV}$ \cite{ATLAS:TOPQ-2012-08}. 
Alternative samples to evaluate systematic uncertainties are generated with
a setting very similar to the one used for $W$/$Z$+jets, using \Alpgen
with up to four additional partons in the ME. 
%
The production of single-top events in the $s$- and $Wt$-channels
is simulated using \Mcatnlo~4.06~\cite{Frixione:2002ik,Frixione:2003ei,Frixione:2005vw} with \Herwig~6.520 showering and
the \texttt{CT10} PDF set, while for the $t$-channel
\Acermc~3.8~\cite{Kersevan:2004} with \Pythia~6.426 showering is used with the \texttt{CTEQ6L1} PDF set. In all samples a top quark mass of \unit[172.5]{GeV} is used consistently.

The SUSY signal samples used in this analysis are generated with
\Pythia~6.426 for the bRPV model and \Herwigpp~2.5.2~\cite{herwig++} for all other
models, with the \texttt{CTEQ6L1} PDF set in all cases.
For all signal models the signal cross sections are calculated
to next-to-leading order in the strong coupling constant,
adding the resummation of soft gluon emission at next-to-leading-logarithmic
accuracy (NLO+NLL)~\cite{Beenakker:1996ch,Kulesza:2008jb,Kulesza:2009kq,Beenakker:2009ha,Beenakker:2011fu}.
The nominal cross section and the uncertainty are taken from an envelope
of cross-section predictions using different PDF sets and factorization
and renormalization scales, as described in \refer~\cite{Kramer:2012bx}.

The decays of tau leptons are simulated directly in the generators
in the case of event samples produced with
\Sherpa, \Herwigpp~2.5.2 and \Pythia~8.165, while in all other cases
\Tauola~2.4~\cite{Jadach:1993tau,Golonka:2006} is used.
For the underlying event model the ATLAS AUET2B tune \cite{ATLAS:AUET2B}
is used for all samples except 
for those generated with \Herwigpp~2.5.2 (UEEE tune \cite{Nadolsky:2008zw}),
with \Pythia 8.165 (AU2 tune \cite{ATLAS:AU2}),
with \Sherpa (which use the built-in \Sherpa tune)
and the \ttbar sample generated with \Powheg (Perugia 2011C tune \cite{Skands:2010}).
All samples are processed either through the {\sc Geant4}-based
simulation of the ATLAS detector~\cite{geant4,Aad:2010wq} or a 
fast simulation framework where showers in the calorimeters are
simulated with a parameterized description \cite{ATLAS:2010af2}
and the rest of the detector is simulated with {\sc Geant4}.
The fast simulation framework is used only for top quark pair production
with \Powheg and the low-\pt $W$/$Z$+jets samples simulated
with \Sherpa. The fast simulation
was validated against full {\sc Geant4} simulation on the
\ttbar sample, where a fraction of the events were simulated in
both frameworks.
In all cases, a realistic treatment of the variation of the number
of $pp$ interactions in the same and neighbouring bunch crossings
(pile-up) is included, with an average of around 20 interactions
per bunch crossing.

For the initial comparison with data, all SM background cross sections
are normalized to  the results of higher-order calculations when available.
The theoretical cross sections for $W$ and $Z$ production are calculated
with DYNNLO~\cite{Catani:2009sm} with the MSTW 2008 NNLO~\cite{Martin:2009iq} PDF set.
The same ratio of the next-to-next-leading-order (NNLO) to leading-order 
cross sections is applied to the production of $W$/$Z$ in association
with heavy-flavour jets. The inclusive \ttbar cross section is calculated
at NNLO, including resummation of next-to-next-to-leading  logarithmic (NNLL)
soft gluon terms, with Top++2.0~\cite{Cacciari:2011hy,Czakon:2011xx} using
MSTW 2008 NNLO PDFs.
Approximate NLO+NNLL calculations are used for single-top production
cross sections~\cite{Kidonakis:2010tc,Kidonakis:2010ux,Kidonakis:2011wy}.
For the diboson sample, the cross section is calculated at NLO with
MCFM~\cite{Campbell:2011bn}, using MSTW 2008 PDFs.

\section{Event reconstruction}
\label{sec:ObjectReco}
%
Vertices consistent with the interaction region and with at least five
associated tracks with $\pt>\unit[400]{MeV}$ are selected; the primary
vertex (PV) is then identified by choosing the vertex with the largest
summed $|\pt|^2$ of the associated tracks~\cite{primvtx}.

Jets are reconstructed from three dimensional calorimeter energy clusters using the anti-$k_t$ jet clustering
algorithm~\cite{Cacciari:2008gp} with distance parameter $R=0.4$. 
Jet momenta are constructed by performing a four-vector sum over clusters of calorimeter cells, treating each as an $(E,\vec{p}\kern0.185em)$ 
four-vector with zero mass.
The jets are corrected for energy from additional pile-up collisions 
using the method suggested in \refer~\cite{Cacciari2008119}, which estimates the
pile-up activity in any given event as well as the sensitivity of any given jet to pile-up. 
Clusters are classified as originating from electromagnetic or
hadronic showers by using the local cluster weighting calibration
method~\cite{Issever2005803}. Based on this classification,
specific energy corrections from a combination of MC simulation
and data~\cite{ATLAS:JES2014} are applied.
A further calibration (jet energy scale) is applied to calibrate
on average the energies of jets to the scale of their constituent
particles~\cite{ATLAS:JES2014}. In this analysis jets are selected
within an acceptance of $|\eta|<2.8$ and are required to have $\pt>\unit[20]{GeV}$.

Jets containing $b$-quarks are used in the analysis to define specific regions where the contribution of background events from $W$/$Z$+jets or \ttbar processes are estimated.
They are identified using a neural-network algorithm~\cite{ATLAS-CONF-2011-102,ATLAS-CONF-2012-043} and a
working point corresponding to 60\% ($<0.5\%$) tagging efficiency for $b$-jets (light-flavour or gluon jets) is used, where the tagging efficiency was studied on 
simulated \ttbar\ events.

Reconstruction of hadronically decaying tau leptons starts from jets with $\pt > \unit[10]{GeV}$
\cite{ATLAS:TauID2013}, and an $\eta$- and $\pt$-dependent energy calibration
to the tau energy scale for hadronic decays is applied \cite{ATLAS:TES2013}.
Discriminating variables based on observables sensitive to the transverse 
and longitudinal shapes of the energy deposits of tau candidates in the calorimeter are combined with tracking information as inputs to a boosted decision tree~(BDT) discriminator.
Measurements from the transition radiation tracker and calorimeter
information are used to veto electrons misidentified as taus. 
Suitable tau lepton candidates must have one or three associated tracks (one or three ``prongs''), with a charge sum of $\pm1$,  and satisfy $\pt>\unit[20]{GeV}$ and $|\eta|<2.5$.
A sample of $Z \rightarrow \tau\tau$ events is used to measure the efficiency of the BDT tau identification. The ``loose'' (``medium'') working points in 
\refer~\cite{ATLAS:TauID2013} are used herein and correspond to an efficiency of approximately 70\% (60\%), independent of \pt, with a rejection 
factor of 10 (20) against jets misidentified as tau candidates (referred to as ``fake'' taus).

Muon candidates are identified by matching one or more track segments in the muon spectrometer~\cite{muonPerf} with an extrapolated inner detector track. 
They are required to have $\pt >$ 10~GeV and $|\eta|<$ 2.4.  
Electron candidates must satisfy $\pt >$ 20~GeV, $|\eta| <$ 2.47 and satisfy the ``Medium++'' identification criteria described in \refer~\cite{elePerf}, re-optimized for 2012 conditions.
Muons and electrons satisfying these identification criteria are referred to as ``baseline'' leptons.

The missing transverse momentum vector \ptmissv and its magnitude, \met, are measured from the transverse momenta of identified jets, electrons, muons and all calorimeter clusters with 
$|\eta| < 4.5$ not associated with such objects~\cite{Aad:2011re}. In the \met measurement tau leptons are not distinguished from jets and it was checked that this does not
introduce a bias in any kinematic variables used in the analysis.

Following object reconstruction, ambiguities between candidate 
jets, taus and light leptons are resolved and further criteria
are applied to select ``signal'' objects. Muons are required
to have $\pt>\GeV{25}$ and to be isolated. The scalar sum of
the transverse momenta of tracks within a cone of size $\Delta R
\equiv \sqrt{(\mathrm{\Delta} \eta)^2 + (\mathrm{\Delta} \phi)^2}= 0.2$
around the muon candidate, excluding the muon candidate track itself,
is  required to be less than \unit[1.8]{GeV}. 
Electrons are required to have $\pt>\GeV{25}$ and pass the ``Tight++'' selection~\cite{elePerf}. The sum of all transverse components of deposits in the 
calorimeter around the electron candidate in a cone of size $\Delta R=0.2$ is required to be less than 10\% of the electron candidate \pt. Finally the electron trajectory is 
required to deviate not more than \unit[1]{mm} in the transverse plane and 
\unit[2]{mm} in the longitudinal direction from the reconstructed PV. Signal jets are required to have $\pt>\GeV{30}$ and to be within the acceptance of the inner detector 
($|\eta| < 2.5$).
Soft central jets ($\pt<\unit[50]{GeV}$, $|\eta|<2.4$) originating from pile-up
collisions are removed by requiring a jet vertex fraction (JVF) above 0.5,
where the JVF is defined as the ratio of the sum of the transverse momentum of
jet-matched tracks that originate from the PV to the sum of
transverse momentum of all tracks associated with the jet.

\section{Event selection}
\label{Sec:EventSelection}
%

\begin{table*}[!tp]
    \caption{Signal region selection criteria for the different channels presented in this paper.
    \label{tab:signalRegions}}\medskip
    {\renewcommand\arraystretch{1.3}
 \begin{tabular}{l|c|c}
  \hline
  \hline
			&\onetau Loose SR&\onetau Tight SR\\
  \hline
Trigger selection&\multicolumn{2}{|c}{$\LJetPt > \GeV{130}$, $\SLJetPt > \GeV{30}$}\\[-0.8ex]
		&\multicolumn{2}{|c}{$\met>\GeV{150}$}\\
		\hline
Taus		&\multicolumn{2}{|c}{$N_\tau^\text{medium}=1$}\\[-0.8ex]
         &\multicolumn{2}{|c}{$\pT>30$\,GeV}\\
        \hline
Light leptons	&\multicolumn{2}{|c}{$N_\ell^\text{baseline}=0\qquad$}\\
		\hline 
Multijet rejection	&\multicolumn{2}{|c}{$ \Delta\phi(\text{jet}_{1,2},\ptmiss) > 0.4 $, $ \Delta \phi(\tau,\ptmiss) > 0.2 $}\\
			
	    \hline
Signal selections	&\multicolumn{2}{|c}{$\mT > 140\GeV$}\\[-0.4ex]
\cline{2-3}
			&$\met > 200\GeV$&$\met > 300\GeV$\\[-0.8ex]
			&$\HT > 800\GeV$&$\HT > 1000\GeV$\\
\hline
\hline

 \end{tabular}
\vspace{40pt}

 \begin{tabular}{l|c|c|c|c}
  \hline
  \hline
			&\twotau Inclusive SR&\twotau GMSB SR&\twotau nGM SR&\twotau bRPV SR\\
  \hline
Trigger selection&\multicolumn{4}{|c}{$\LJetPt > \GeV{130}$, $\SLJetPt > \GeV{30}$}\\[-1.0ex]
		&\multicolumn{4}{|c}{$\met>\GeV{150}$}\\
		\hline
Taus		&\multicolumn{4}{|c}{$N_\tau^\text{loose}\geq2$}\\[-1.0ex]
         &\multicolumn{4}{|c}{$\pT>20$\,GeV} \\
        \hline
Light leptons	&\multicolumn{4}{|c}{$N_\ell^\text{baseline}=0\qquad$}\\
		\hline 
Multijet rejection$\!$	&\multicolumn{4}{|c}{$\Delta\phi(\text{jet}_{1,2},\ptmiss)\geq 0.3$}\\
			
	    \hline
Signal selections	&$\mTTkerned\!\geq\!\unit[150]{GeV}\!$&\multicolumn{2}{|c|}{$\mTT\geq\unit[250]{GeV}$}&$\mTTkerned\!\geq\!\unit[150]{GeV}\!$\\
\cline{2-5}
			&$\HTtj > 1000\GeV$&$\HTtj > 1000\GeV$&$\HTtj > 600\GeV$&$\HTtj > 1000\GeV$\\[-1.1ex]
			&&$\njet\geq4$&$\njet\geq4$&$\njet\geq4$\\
\hline
\hline

 \end{tabular}
\vspace{40pt}

 \begin{tabular}{l|c|c|c|c}
  \hline
  \hline
  &\tauleps GMSB SR&\tauleps nGM SR & \tauleps bRPV SR & \tauleps mSUGRA SR\\
  \hline
  Trigger selection&\multicolumn{4}{|c}{$p_\text{T}^\ell>\GeV{25}$}\\
  \hline
  Taus &\multicolumn{4}{|c}{$N_\tau^\text{loose}\geq1$}\\[-1.0ex]
  & \multicolumn{4}{|c}{$\pT>20$\,GeV} \\
  \hline
  Light leptons	&\multicolumn{4}{|c}{$N_\ell^\text{signal}=1,\quad N_\text{other lep}^\text{baseline}=0$}\\
  \hline 
  Multijet rejection 	&\multicolumn{4}{|c}{$m_\text{T}^\ell >$ 100 GeV}\\
\hline
   Signal selections & $\meff > \GeV{1700}$ & $\met >\GeV{350}$&$\meff>\GeV{1300}$& $\met >\GeV{300}$\\[-1.1ex]
   & & $N_\text{jet}\geq3$&$N_\text{jet}\geq4$&$N_\text{jet}\geq3$\\

  \hline
  \hline

 \end{tabular}
 }

\end{table*}
For the \onetau channel, events with only one hadronically decaying
medium tau lepton candidate with $\pT>\unit[30]{GeV}$, no additional
loose tau candidates, and no candidate muons or electrons are selected;
in the \twotau channel, events are selected with two or more loose tau
leptons with $\pT>\unit[20]{GeV}$ and no candidate muons or electrons;
events in the \tauel and \taumu channels have one or more loose tau
candidates with $\pT>\unit[20]{GeV}$ and one additional signal electron
or muon, respectively.

All events have to fulfil a common initial set of requirements, 
in the following referred to as the ``preselection''.
Events are required to have a reconstructed PV, to have no jets or muons
that show signs of problematic reconstruction, to have no jets failing
to satisfy quality criteria, and to have no muons that are likely to have originated from
cosmic rays.

After the preselection, several requirements are applied to
define various signal regions (SRs) in each final state. 
The individual SRs have been optimized for specific signal
models and are combined in the final results for the respective
signal scenarios. Two SRs (\onetau ``Loose'' and \twotau 
``Inclusive'') are designed with relaxed selections
to maintain sensitivity for other BSM scenarios and to provide
model independent limits.

The following variables are used to suppress the main background processes ($W$+jets, $Z$+jets
and top, including \ttbar\ and single-top events) in each final state:
\begin{itemize}

\item \mT{}, the transverse mass formed by \met and the \pt ~of
  the tau lepton in the \onetau channel \\
  $\mT=\sqrt{2\pt^{\tau}\met(1-\cos(\Delta\phi(\tau,\ptmiss)))}$.
  In addition the variable \mTT is used as a discriminating
  variable in the \twotau channel;

\item $m_\text{T}^{\ell}$, the transverse mass formed by \met and the \pt ~of
  the light leptons\\
   $\mtlep=\sqrt{2\pt^{\ell}\met(1-\cos(\Delta\phi(\ell,\ptmiss)))}$~;

\item \HT{}, the scalar sum of the transverse momenta of the tau, light
    lepton and signal jet ($\pt>\unit[30]{GeV}$) candidates in the event: \\%
    $\HT=\sum_{\text{all}\;\ell}\pt^\ell+\sum_{\text{all}\;\tau}\pt^\tau
     + \sum_{\text{all jets}} \pt^\mathrm{jet}$~;

\item \HTtj{}, the scalar sum of the transverse momenta of the tau and light lepton candidates
    and the two jets with the largest transverse momenta in the event:\\%
    $\HTtj=\sum_{\text{all}\;\ell}\pt^\ell+\sum_{\text{all}\;\tau}\pt^\tau
       + \sum_{i=1,2} \pt^{\mathrm{jet}_i}$~;

\item the magnitude of the missing transverse momentum \met;

\item the effective mass $\meff=\HTtj+\met$;

\item the number of reconstructed signal jets \njet.

\end{itemize}

While optimizing the choice of variables, studies showed that there is a correlation between \HT{} and \njet, 
given that the sum of the jet \pT{} is used in the defintion of \HT{}.
In the \twotau and \taulep channels, where a selection on \njet is used to define different SRs, the variable \HTtj 
is used in order to avoid such correlation.

\subsection*{\texorpdfstring{\onetau}{One-tau} signal regions}

The various selection criteria used to define the two SRs in the \onetau channel are summarized in \tab~\ref{tab:signalRegions}.
A requirement on the azimuthal angle between \ptmissv and either of the two leading jets ($\Delta\phi(\text{jet}_{1,2},\ptmiss)$) is used to remove
multijet events, where the \met arises from mismeasured highly energetic jets. 
To further reduce these events in the SRs, a tighter selection on \met is also applied.
The transverse mass \mT{} is used to remove $W$+jets events, 
while a requirement on \HT{} is applied in order to reduce the 
contribution of all remaining backgrounds.

The main SR (``tight SR'') applies tight selections on \met and \HT{} as a result of optimizing the sensitivity in the high-$\Lambda$ region of the GMSB 
model parameter space, given that lower mass regions were excluded in earlier analyses.
A ``loose SR'', with looser requirements on \met and \HT{}, is also defined and used to calculate model-independent limits. 
In the GMSB model the strong production cross section, for which the analysis has the largest sensitivity,
decreases faster with increasing $\Lambda$ than the cross sections
for weak production. Therefore, the selection efficiency with
respect to the total SUSY production decreases for large values of $\Lambda$.
For high $\tan\beta$, the product of acceptance and efficiency is of the order of 0.3\%, decreasing to 0.1\% for low $\tan\beta$.
The tight SR yields the best sensitivity in the high-\moh, low-\mz region
of the mSUGRA and bRPV models and, 
when combined with the other channels,
extends the overall sensitivity range in these models.
In the mSUGRA model the product of acceptance and efficiency for the tight
signal selection ranges from the permille level to around 4\%,
with the higher values being observed in the low $m_{1/2}$ region.
In the bRPV signal region the product of acceptance and efficiency for the tight SR ranges from the permille level
to around 1\% (tight SR), with the higher values being observed in the low-\mz, high-\moh region.
The \onetau channel does not contribute to the nGM scenario where by construction each event contains at least two high-\pt taus.

\subsection*{\texorpdfstring{\twotau}{Two-tau} signal regions}

The criteria used to define the four SRs in the \twotau channel are shown in \tab~\ref{tab:signalRegions}. 
Multijet events are rejected by a requirement on $\Delta\phi(\text{jet}_{1,2},\ptmiss)$,
while $Z$+jets events are efficiently removed by a requirement on \mTT{}.
A selection on \HTtj{} is then applied in order to reduce the contribution of all remaining backgrounds.
Additional requirements on the number of jets in the event are also used to define SRs that are sensitive in specific signal models.

The GMSB SR was optimized to be sensitive to the high-$\Lambda$ region of the parameter space.
For high $\tan\beta$ the product of acceptance and efficiency is of the order of 0.5\%, falling to 0.2\% for low $\tan\beta$.
The nGM SR was optimized for high gluino masses. Given the topology of the signal events, at least four jets are required and a lower requirement
on the value of \HTtj{} with respect to the GMSB SR is applied. 
In this model the gluino pair production cross section is primarily a function of $m_{\tilde{g}}$, ranging from $\unit[17.2]{pb}$ for $m_{\tilde{g}}=400$~GeV to
$\unit[7]{fb}$ for $m_{\tilde{g}}=1100$~GeV. The product of acceptance and efficiency for this channel in the nGM model is of the order of 4\% for high $m_{\tilde{g}}$, 
independent of $m_{\tilde{\tau}}$, and it falls to $\sim$2\% for low $m_{\tilde{g}}$ due to the analysis requirements on the \pT{} of the leading jet and on \met.
The \twotau channel has extremely small acceptance in the mSUGRA model, due to the requirement of a second high-\pt ~tau; for this reason no SR optimized for this
scenario is defined. 
In the bRPV SR the selection was optimized to be sensitive in the low-\mz, high-\moh region of the parameter space, where the branching ratio to events with two 
real taus is highest.
The product of acceptance and efficiency of the dedicated SR is of the order of 1\% in the most sensitive regions of the parameter space, decreasing to the permille level 
in other regions.

\subsection*{\texorpdfstring{\taulep}{Tau+Lepton} signal regions}

Events from multijet production and from decays of $W$ bosons 
into a light lepton and a neutrino, which constitute the largest source of SM background,
are suppressed by requiring $\mtlep >$ \unit[100]{GeV}. Different SRs are then
defined by applying further requirements on
\met, \meff and \njet to yield good sensitivity to each of the considered signal models.
In the GMSB model, the SR  selection was also optimized for the high-$\Lambda$ region; a tight requirement
on \meff is applied to significantly reduce the contribution of all backgrounds.
The product of acceptance and efficiency in this SR varies between 0.2\% to 0.4\% across the ($\Lambda$, $\tan\beta$) plane.
The nGM SR was optimized for high gluino masses. Since a high jet multiplicity is expected in this scenario, events with at least three signal jets are selected.
The remaining background contribution is reduced with a requirement on \met.
The product of acceptance and efficiency of this selection is of the order of 2\% for high $m_{\tilde{g}}$, decreasing to 0.2\% for lower values of the gluino mass.
Requirements similar to those for the nGM SR are applied to define the mSUGRA SR, which was optimized to be sensitive in a low-\moh and high-\mz region of the parameter space.
The product of acceptance and efficiency in this case ranges from the permille level to 2\% across the parameter space.
For the bRPV SR the selection optimization is performed in a high-\mz,
medium-\moh region of the parameter space.
At least four signal jets are required and the remaining background
contribution is reduced with a requirement on $\meff$.
The product of acceptance and efficiency also in this case ranges from the permille level to 2\%.
The full list of criteria used to define the different SRs in the \tauel and \taumu channels is given in \tab~\ref{tab:signalRegions}.

\FloatBarrier

\section{Background estimation}
\label{sec:BgEstimation}
%
The background in this analysis arises predominantly from $W$+jets,
$Z$+jets, top and multijet events, with contributions from ``true'' taus
and ``fake'' taus (jets misidentified as taus).
The contributions of these backgrounds in the various signal regions
are estimated from data.
Because of the differences of the topologies in the four final states considered, different
techniques are employed to estimate the multijet background.
\Tab~\ref{tab:bgMethods} gives an overview of all the different methods used
for the background estimation in all channels, which are described in the following subsections. 
The small diboson background
contributions are estimated using MC simulations, while the contributions from
other backgrounds like low mass Drell--Yan,
\ttbar+$V$ and $H\to\tau\tau$ were found to be negligible.

\begin{table*}[!tp]
\centering
\caption{Overview of the various techniques employed
for background estimation.}\medskip
\label{tab:bgMethods}

\begin{tabular}{l|c|c|c}
\hline
\hline
Background & \onetau & \twotau & \taulep\\
\hline
$W$+jets (true) & matrix inversion & \multirow{2}{*}{matrix inversion} & --\\
$W$+jets (fake) & matrix inversion & & matrix inversion \\
\hline
$Z$+jets (true) & with $W$+jets & matrix inversion & --\\
$Z$+jets (fake) & with $W$+jets & -- & --\\
\hline
Top (true) & matrix inversion &\multirow{2}{*}{matrix inversion}  & matrix inversion\\
Top (fake) & matrix inversion & & matrix inversion\\
\hline
Multijets & ABCD method & jet-smearing method & matrix method\\
\hline
Dibosons & from simulation & from simulation & from simulation\\
\hline
\hline
\end{tabular}
\end{table*}

\begin{table*}[!tp]
\centering
\caption{Overview of the various control regions employed
for the background estimation of $W$, $Z$ and top quark backgrounds. 
Trigger requirements and selected objects are identical to the signal region requirements in the respective channels.}
\label{tab:EWCRs}

\vspace{10pt}
 \subtable[Control region selections in the \onetau analysis. A multijet rejection cut $ \Delta\phi(\text{jet}_{1,2},\ptmiss) > 0.4 $ is applied in all CRs.]{
 \renewcommand\arraystretch{1.15}
 \begin{tabular}{c|c|c}
 \hline
 \hline
    & $\nbjet=0$ & $\nbjet>0$ \\
    \hline    
    $\mT < \unit[90]{GeV}$ & $\mathrm{CR_{WTrue}}$ & $\mathrm{CR_{TTrue}}$\\[-0.5ex]
    or $\Delta\phi(\tau,\ptmiss)<1.0$ & & \\[-0.5ex]
    or $\pt^\tau>\unit[55]{GeV}$ & & \\
    \hline
    $\unit[90]{GeV}<\mT<\unit[140]{GeV}$ & $\mathrm{CR_{WFake}}$ & $\mathrm{CR_{TFake}}$ \\[-0.5ex]
    and $\Delta\phi(\tau,\ptmiss)>1.0$ & & \\[-0.5ex]
    and $\pt^\tau<\unit[55]{GeV}$ & & \\
 \hline
 \hline
  \end{tabular}}
 
 \vspace{10pt}
\subtable[Control region selections in the \twotau analysis. A multijet rejection cut $ \Delta\phi(\text{jet}_{1,2},\ptmiss) > 0.3 $ is applied in all CRs.]{
 \renewcommand\arraystretch{1.15}
 \begin{tabular}{c|c|c}
 \hline
 \hline
  Top CR	& $W$ CR		& $Z$ CR	\\
 \hline
 \multicolumn{3}{c}{$\HTtj<\unit[550]{GeV}$}	\\
 \hline
 \multicolumn{2}{c|}{$\mTT> \unit[150/200]{GeV}$}	& $\mTT< \unit[80]{GeV}$	\\
 \cline{1-2}
 $\nbjet>0$ & $\nbjet = 0$ &--- \\
 \hline
 \hline
 \end{tabular}
 }
 
\vspace{10pt}
\subtable[Control region selections in the \taulep analysis.]{
 \renewcommand\arraystretch{1.15}
 \begin{tabular}{c|c|c}
 \hline
  \hline
  Top fake-tau CR & Top true-tau CR & {$W$ CR} \\ 
  \hline
  \multicolumn{3}{c}{50 GeV $<  \met <$ 130 GeV} \\[-0.5ex]
  \multicolumn{3}{c}{50 GeV $< \mtlep <$ 190 GeV} \\[-0.5ex]
  \multicolumn{3}{c}{$\meff <$ 1000 GeV}\\
 \hline
  \multicolumn{2}{c|}{$\nbjet \geq1$} & $\nbjet = 0$\\ 
  \cline{1-2}
  50 GeV $< \mtlep <$ 120 GeV& 120 GeV $< \mtlep <$ 190 GeV & \\ 

  \hline
 \hline
 \end{tabular}}
\end{table*}

\subsection{$W$, $Z$ and top quark backgrounds}

The main estimation technique for electroweak and top quark backgrounds is referred to in the following as the ``matrix inversion'' method. 
In each signal region, the SM background predicted by MC simulation
is scaled by factors obtained from appropriately defined control regions (CRs).
This is done to reduce the impact of possible mis-modelling of
tau misidentification probabilities and kinematics 
in the MC simulations. The CRs are chosen such that:

\begin{itemize}
 \item they are as kinematically close as possible to the final signal regions, without overlapping with them, while having low signal contamination;
 \item each CR is enriched with a specific background process;
 \item the tau misidentification probability is, to a good approximation, independent of the kinematic variables used to separate the SR from the CRs.
\end{itemize}

By doing this, the measured ratio of the data to MC event yields in the CR can be used to compute scaling factors to correct the MC background prediction in the SR.
The vector defined by the scaling factors for each background ($\vec{\omega}$) is obtained by inverting the equation 
$\vec{N}^{\text{data}}=A\,\vec{\omega}$, where 
$\vec{N}^{\text{data\,}}$ is the observed number of data events in each CR, after subtracting the expected number of events from other SM processes,
and the matrix $A$ is obtained from the MC expectation for the number of events originating from each of the backgrounds ($W$, $Z$ and top). 
The signal contamination in all CRs has been determined from MC simulation and is
well below 5\%, except for the nGM SR in the \twotau channel where 
up to 10\% contamination is observed.%
\footnote{It was checked that this contamination has a negligible effect
          on the limit obtained in this scenario.}
Correlations due to the contribution of each background process in the different CRs are properly taken into account in the matrix $A$.
To obtain the statistical uncertainties on the scaling factors, all contributing parameters are varied within their uncertainties, the procedure is repeated and new scaling factors are obtained.  The width of the distribution of the resulting scaling factors is then used as their statistical uncertainty. 

\subsubsection*{\texorpdfstring{\onetau}{One-tau} channel}

The dominant backgrounds to the \onetau SR arise from $W$+jets,
$Z$+jets and \ttbar. Events can be divided into those which
contain a true tau and those in which a jet is misidentified as a tau. 
Since the composition of true and fake taus in the control region and signal region may differ, it is necessary 
to compute separate scaling factors for events with true and fake taus. For this purpose, the CRs are defined by using two variables: the transverse mass, used to separate true and fake 
taus, and the $b$-tagging, used to provide a top-enriched (\ttbar CR) or top-depleted ($W$ or $Z$ CR) sample. 
The contribution in these CRs from other backgrounds (e.g. multijet background) is negligible.
The full list of selection requirements for these control regions, after the 
preselection, tau selection and light-lepton veto requirements are applied, is provided in \tab~\ref{tab:EWCRs}. The matrix $A$ is a $4\times4$ matrix from which the scale factors for $W$ 
events with a true tau candidate, $W/Z$ events with a fake tau candidate, and top events with either a true or a fake tau candidate are obtained. 
In $Z$+jets events, the background is dominated by $Z$ decays to neutrinos, and therefore the tau candidate is typically a misidentified jet. For this reason, 
the scaling factor is obtained from the CR defined for $W$+jets (fake) events. 

\begin{figure}[!tp]
  \subfigure[$N_\mathrm{jets}$ distribution in inclusive $W$/$Z$ and \ttbar region]
            {\includegraphics[width=0.51\textwidth]{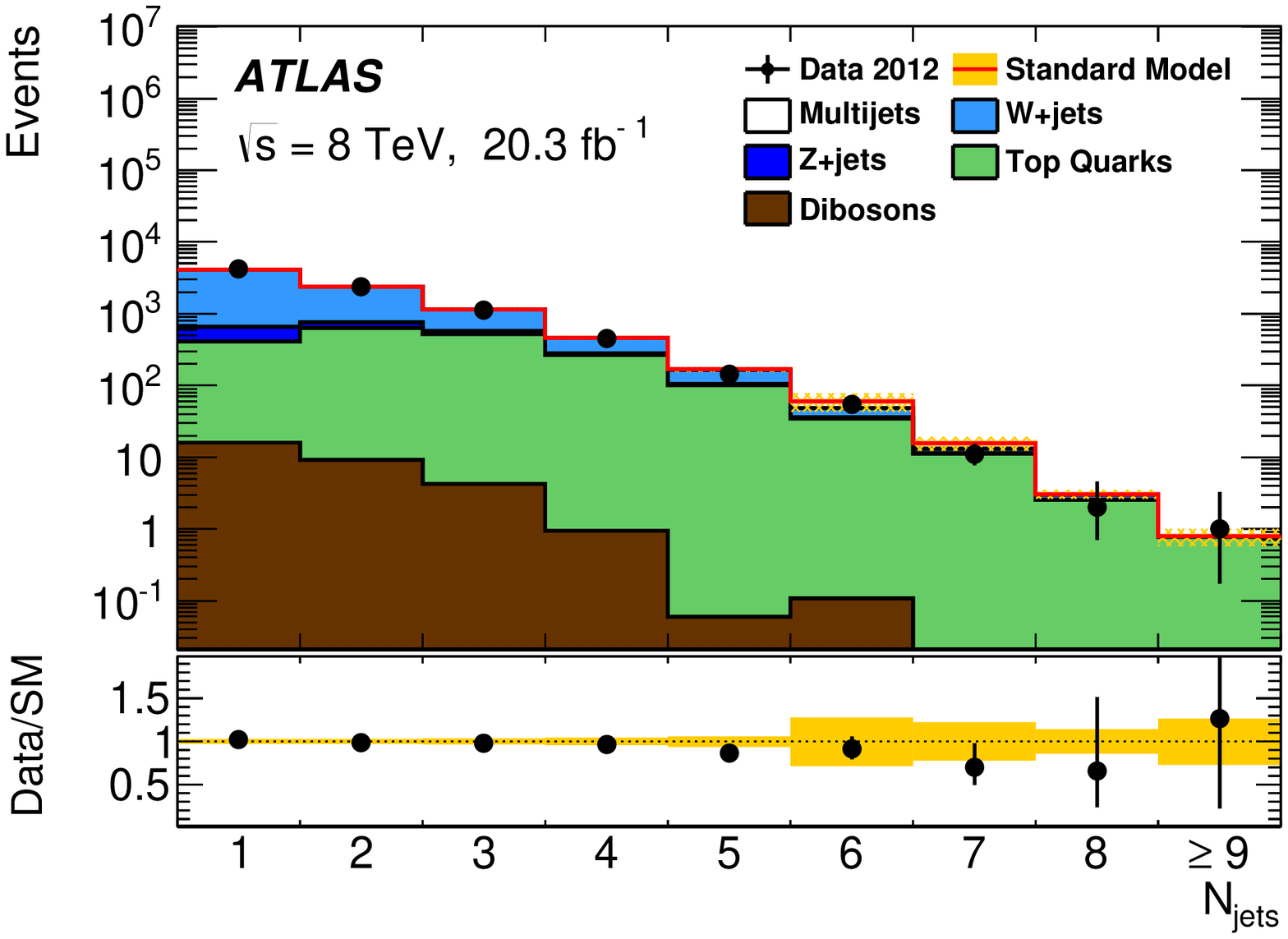}}
  \subfigure[\mT{} distribution in the \ttbar validation region]
            {\includegraphics[width=0.51\textwidth]{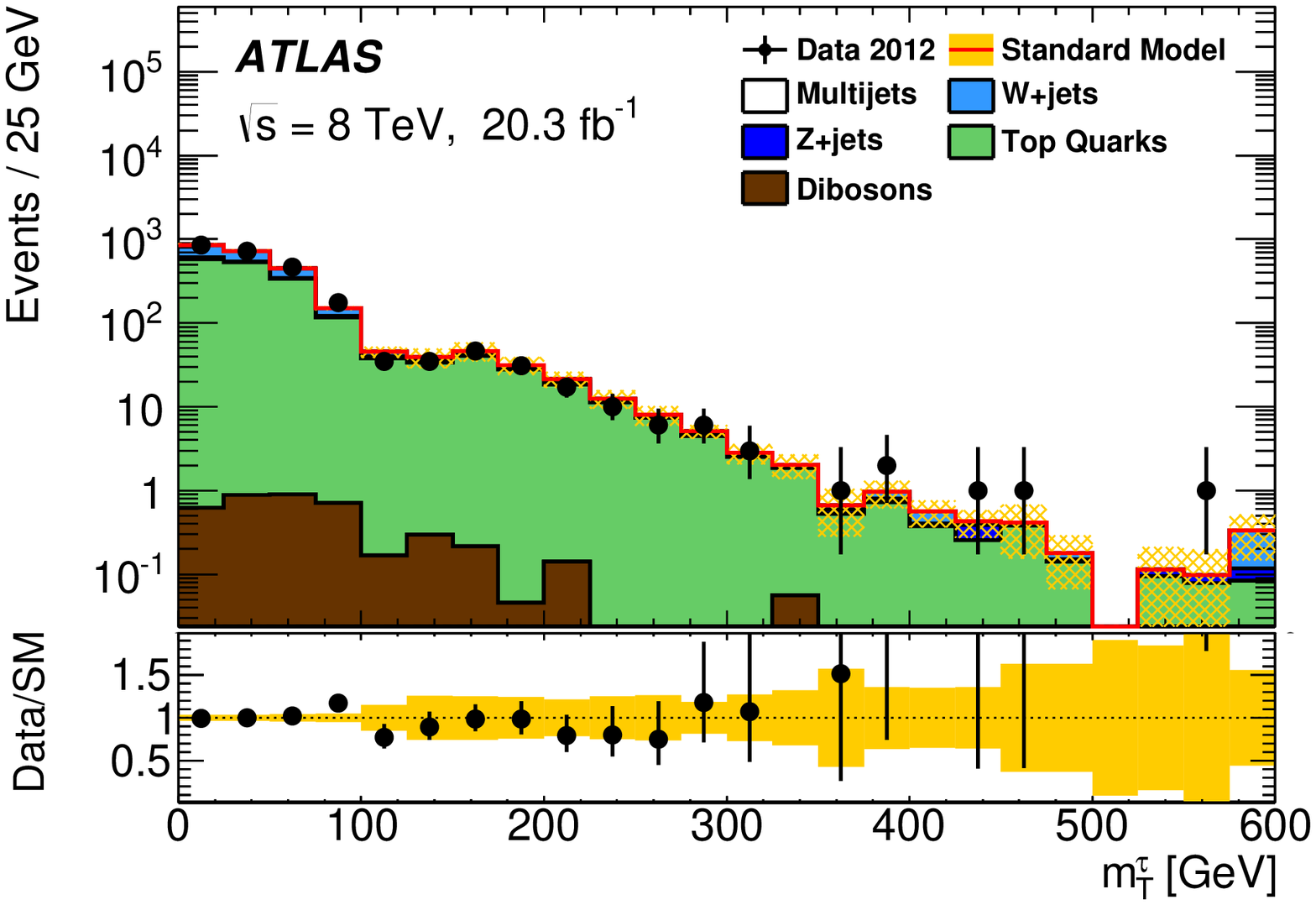}}
  \caption{Kinematic distributions in the \onetau channel for events (a) in an
           inclusive $W$/$Z$ and \ttbar validation region and (b)
           \ttbar-enriched validation region. Data are represented by the points.
           All backgrounds are scaled according to the results of
           the data-driven background estimates.
           The shaded band centred around the total background indicates
           the statistical uncertainty on the background expectation.
\label{fig:valRegion1tau}}
\end{figure}%
Typical scaling factors obtained for the various MC samples are $\sim$\unit[0.6] for $W$+jets, $Z$+jets and $\sim$\unit[1.0] for \ttbar with fake taus, while they are
$\sim$\unit[1.1] for $W$+jets and $\sim$\unit[1.0] for \ttbar with true taus. 
The comparatively large scale factor for $W$+jets and $Z$+jets with fake tau candidates
reflects the insufficient description in MC simulation of narrow jets, which in these events are
predominantly initiated by colour-connected light quarks, as opposed to the fake tau candidates in ttbar events.
The associated statistical uncertainties on these scaling factors are in the range of 5--50\%, 
depending on the CR. Good agreement between data and scaled MC events is observed in the relevant kinematic distributions in the CRs. 
\Fig~\ref{fig:valRegion1tau}(a) shows the jet multiplicity distribution (an independent variable not used for background separation) 
on an inclusive data sample made from the four CRs, extending the kinematic range up to (but excluding) the SR.
A \ttbar-enriched validation region is formed from the inclusive
sample by means of $b$-tagging, and the corresponding \mT distribution is
shown in \fig~\ref{fig:valRegion1tau}(b). It shows good agreement in the
true-tau-dominated low-\mT range as well as for $\mT>\unit[140]{GeV}$ (beyond the CR),
where events with either a true or a fake tau candidate contribute with similar amounts.

\subsubsection*{\texorpdfstring{\twotau}{Two-tau} channel}
In the \twotau analysis, the backgrounds from $W$+jets and \ttbar are dominated by events in which one tau candidate is a true tau and the other is a jet misidentified as a tau. 
The contributions from $Z$+jets events are dominated by final states with $Z\to\tau\tau$ decays. The definitions of the \twotau control regions are given in \tab~\ref{tab:EWCRs}. Three CRs are 
defined, for $W$+jets, $Z$+jets and \ttbar ~events. All CRs have a negligible contamination from multijet events due to the requirement on $\Delta\phi(\text{jet}_{1,2},\ptmiss)$. 
Given that the ratio of true to fake tau candidates in the CR and SR is the same, as confirmed by generator-level MC studies, there is no need to separate the
CRs for fake tau and true tau backgrounds. The matrix $A$ in this case is a $3\times3$ matrix from which the scale factors for $W$, $Z$ and top events are obtained. The selection criteria
$\mTT>\GeV{150}$ (for the Inclusive and bRPV SR) or $\mTT>\GeV{200}$ (for the GMSB and nGM SR) are applied to reproduce the signal region kinematics.

\begin{figure}[!tp]
  \subfigure[Tau \pt distribution in combined $W$ and \ttbar CR]{\includegraphics[width=0.51\textwidth]{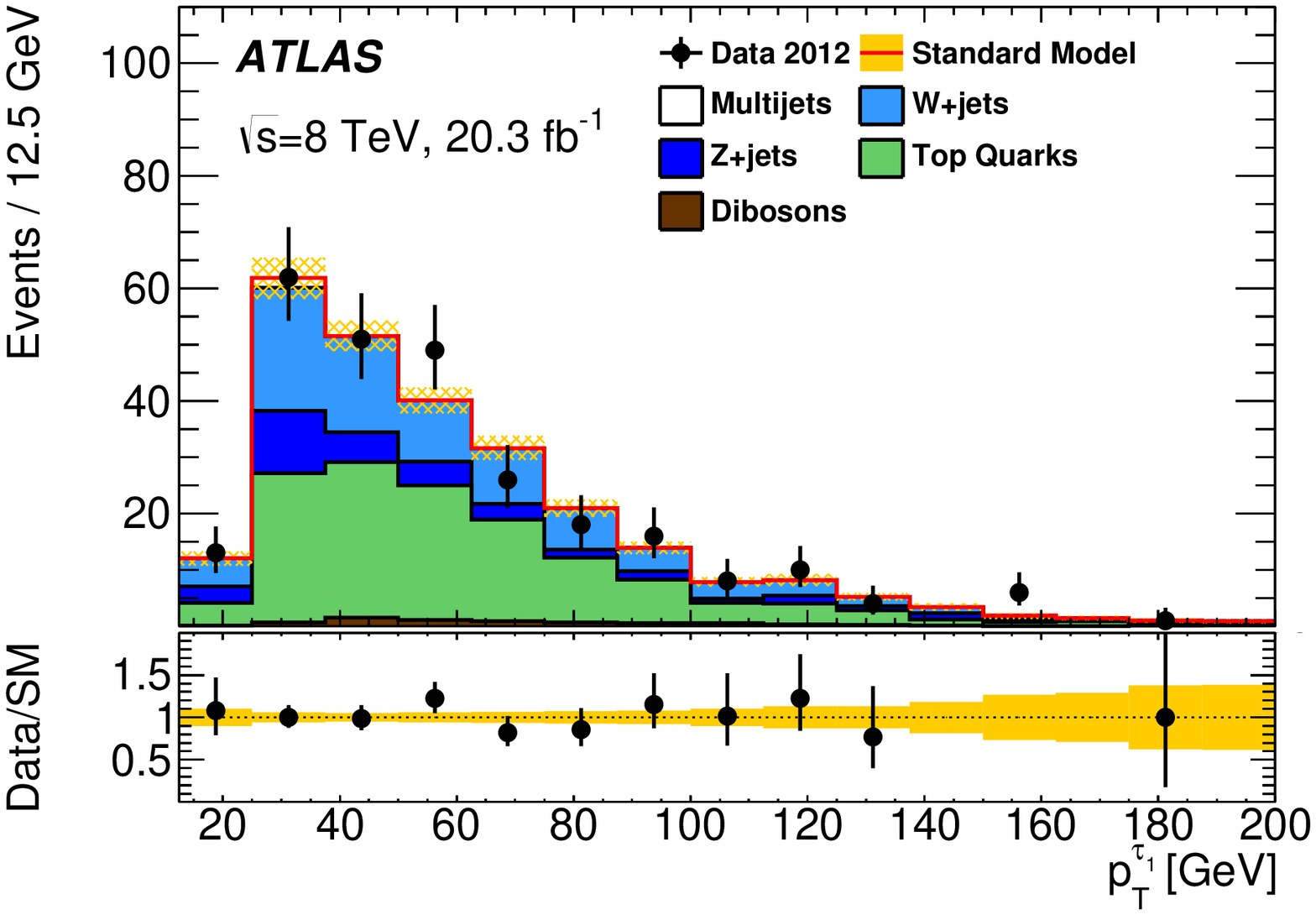}}
  \subfigure[\mTone{} distribution in the multijet VR]{\includegraphics[width=0.51\textwidth]{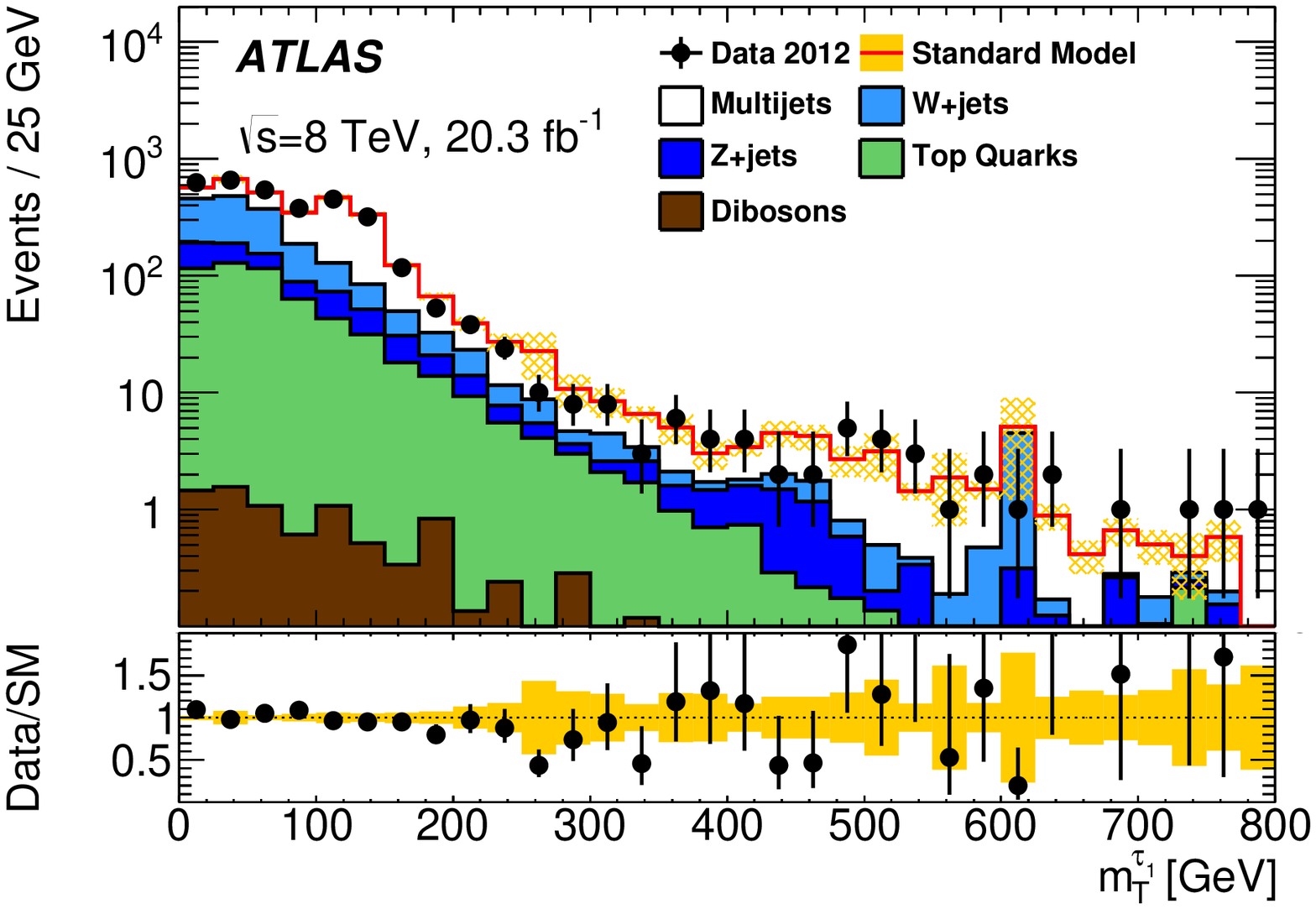}}
\caption{Kinematic distributions for events (a) in the \twotau $W$ and
         \ttbar control region and (b) in the multijet validation region.
         Data are represented by the points. All backgrounds are
         scaled according to the results of the data-driven background
         estimates. The shaded band centred around the total background
         indicates the statistical uncertainty on the background expectation.
\label{fig:valRegion2tau}}
\end{figure}

Typical scaling factors obtained for various MC samples are $\sim$\unit[0.6]{} for the $W$+jets, $\sim$\unit[1.4]{} for the $Z$+jets and $\sim$\unit[0.9]{} for \ttbar, with associated statistical 
uncertainties in the range of 10--30\%.
Good agreement between data and scaled MC events in the relevant kinematic
distributions is observed in the CRs. An example can be seen in
\fig~\ref{fig:valRegion2tau}(a), where the distribution of the
transverse momentum of the leading tau candidate in data and scaled MC
is compared in an inclusive CR defined by combining the $W$
and \ttbar CRs discussed in this section.

\subsubsection*{\texorpdfstring{\taulep}{Tau+lepton} channel}

In the \taulep analysis the ratio of real to fake taus depends on the background process. For $W$ decays, due to the high efficiency and purity of the electron and muon reconstruction, 
the light lepton is always a real lepton from the $W$ decay, while the tau is faked by a recoiling hadronic object.
For \ttbar the light lepton originates from the decay chain of one of the top quarks, while the tau can either be a real tau from the decay of the other top or a fake tau from 
a jet in the event. $Z$ decays do not contribute a significant amount to the background and are estimated from simulation.

Three control regions are defined for $W$, \ttbar{} with fake taus and \ttbar{} with true taus. Events with true or fake taus are separated by using a requirement on the \mtlep of the event, 
as summarized in \tab~\ref{tab:EWCRs}.
The matrix $A$ in this case is a $3\times3$ matrix from which the scale factors for $W$, top with true taus and top with fake taus are obtained.

\begin{figure}[ht]
  \subfigure[$m_\text{T}^\ell$ distribution (\tauel)]{\includegraphics[width=0.51\textwidth]{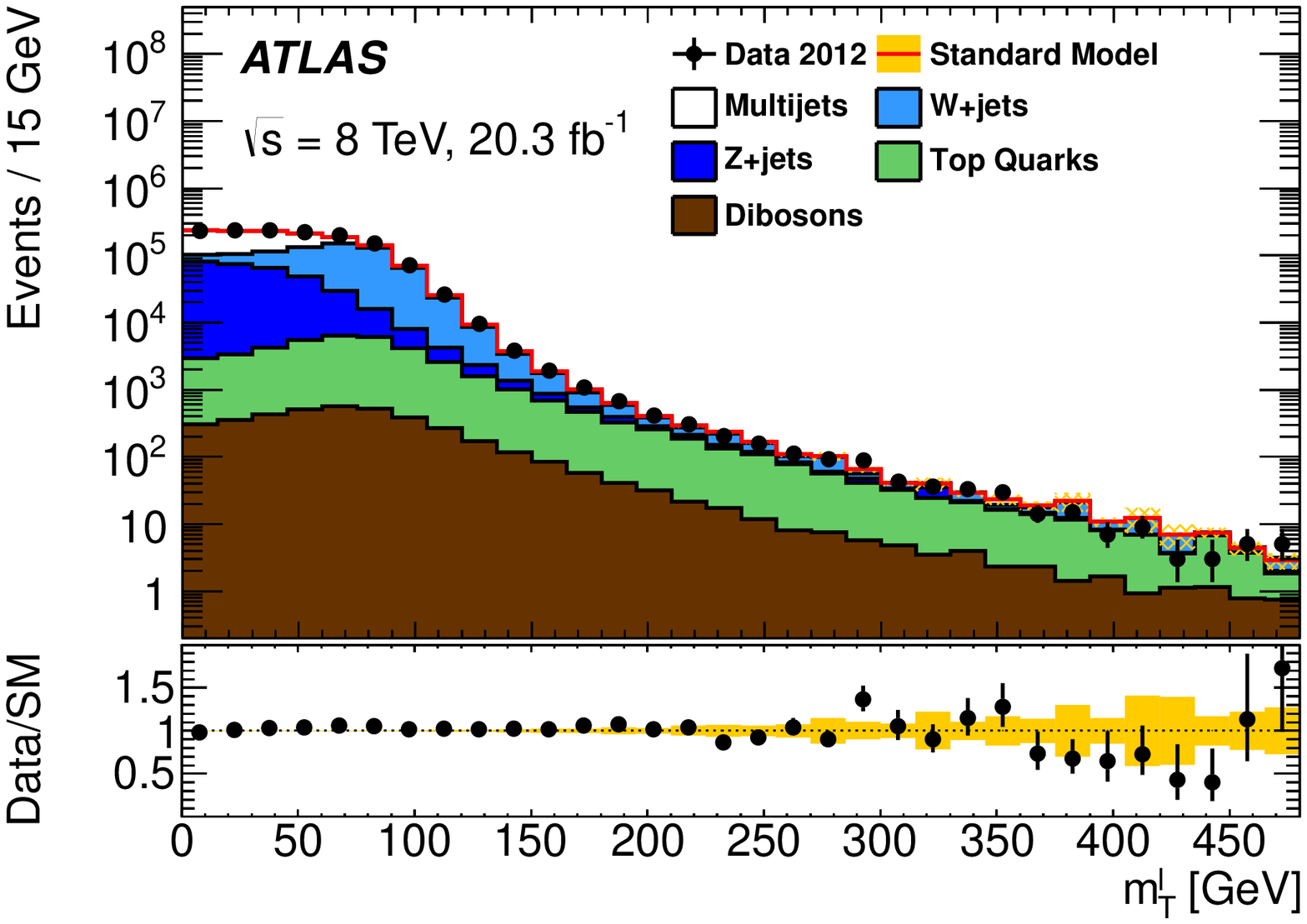}}
  \subfigure[$m_\text{T}^\ell$ distribution (\taumu)]{\includegraphics[width=0.51\textwidth]{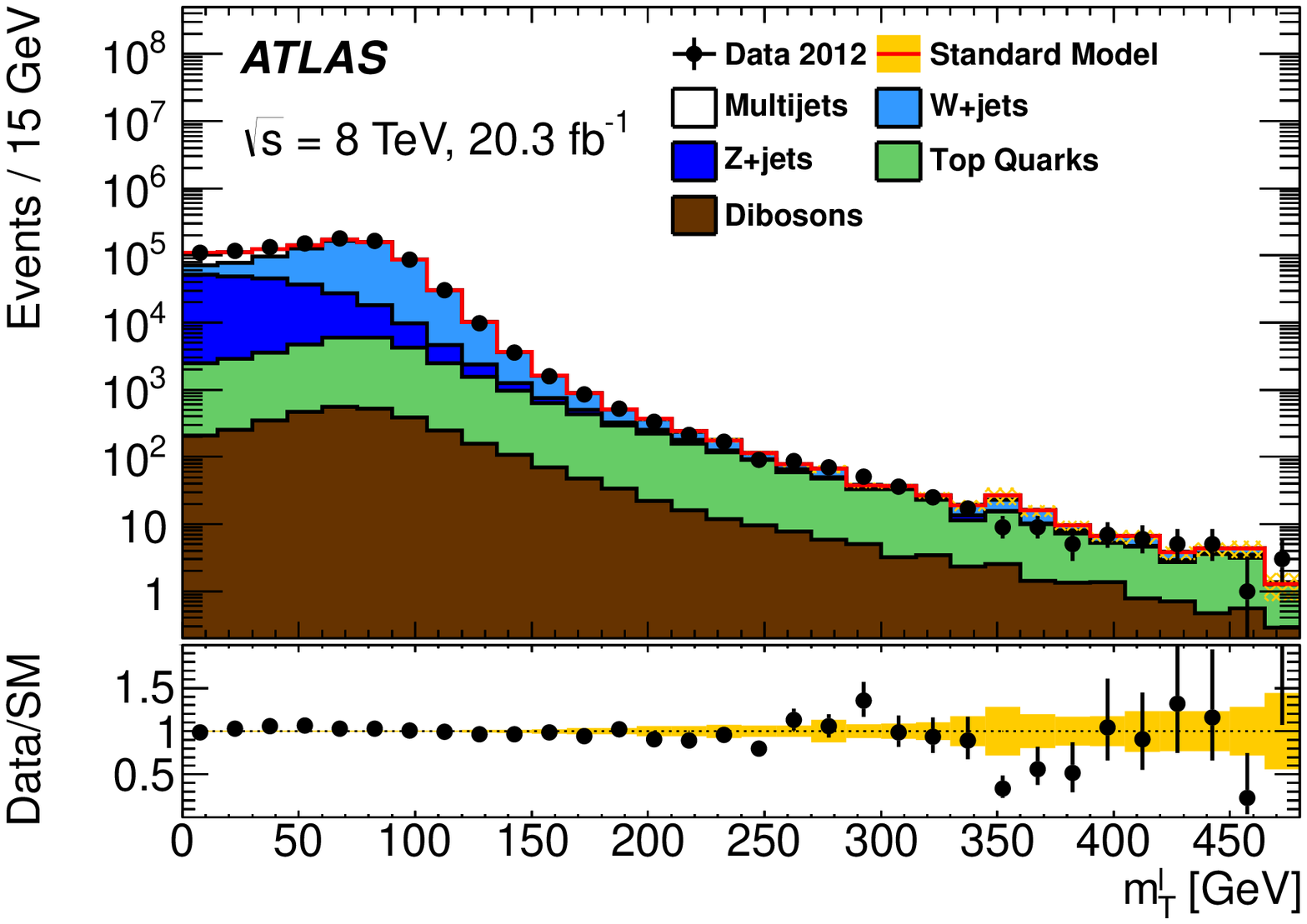}}
  \caption{Kinematic distributions in the \taulep combined W and \ttbar
           control regions. Data are represented by the points.
           All backgrounds are scaled according to the results
           of the data-driven background estimates and the multijet background is estimated as described in \sect~\ref{sec:multijet}.
           The shaded band centred around the total background
           indicates the statistical uncertainty on the background expectation.
\label{fig:valRegiontaulep}}
\end{figure}

Typical scaling factors obtained are $\sim$\unit[0.7]{} for the $W$+jets, $\sim$\unit[0.9]{} for the \ttbar{} with a fake tau and $\sim$\unit[0.8]{} for \ttbar{} 
with a true tau. The associated statistical uncertainties are of the order of 20\%.
An example of the very good agreement in the CRs between data and scaled MC 
is shown in \fig~\ref{fig:valRegiontaulep}, which presents the \mtlep{} distribution
for the \tauel and \taumu channels in a combined $W$ and \ttbar CR defined as the
CR selection apart from the cut on the variable plotted.

\subsection{Multijet backgrounds}
\label{sec:multijet}

To estimate the multijet background contribution in the signal regions, different methods are employed for each of the three channels.

\subsubsection*{\texorpdfstring{\onetau}{One-tau} channel}

For the \onetau channel, the contribution arising from multijet background
processes due to fake taus is estimated from data using the so-called
``ABCD'' method. Four exclusive regions, labelled A, B, C and D, are
defined in a two-dimensional plane specified by two discriminating
variables that are uncorrelated for background
events: the tau identification tightness and a combination of \met and
its angular separation in $\phi$ to either of the leading and sub-leading
jets (\tab~\ref{tab:QCDCRs}). 
To increase the number of events in regions A and C, very loose tau candidates are defined by taking the nominal (medium) tau selection and relaxing the criteria on the 
BDT discriminant. 
Region D is defined to be similar to the SR, except for the fact that the requirement on \met is inverted and there is no requirement on \HT.
Multijet events in region D may be estimated because the ratio of the numbers of events in regions A and B is equal to the ratio of numbers of events in regions C and D.
Therefore, the number of events in region D ($N_\mathrm{D}$) is $N_\mathrm{D} = c \times N_\mathrm{B}$, where $N_\mathrm{B}$ is the number of events in region B and $c = \frac{N_\mathrm{C}}{N_\mathrm{A}}$ is the ``correction factor''.
In order to estimate the total yield from multijet events in the final SR, the number of events 
obtained in region D is scaled by the fraction of events passing the final requirements on \HT{} and \met.
This fraction is derived in region A, after checking that it has little dependence on the requirements used to define the different multijet regions.
In each region, the non-multijet contribution is estimated using MC events scaled according to the procedure detailed in the previous section, and is subtracted from the data. 

\begin{table}[!tp]
  \centering
  \caption{Definitions of control regions used in the estimates of the multijet backgrounds.}
  \label{tab:QCDCRs}
\smallskip
  \subtable[Regions used in the ABCD method for the \onetau analysis.
           The requirement on \HT{} is not applied in the definition of these control regions.\smallskip]{
 \renewcommand\arraystretch{1.12}
 \begin{tabular}{c|cc}
        \hline
        \hline
        & Very loose tau & Nominal tau \\
    \hline
        $\begin{array}{l}\Delta\phi(\text{jet}_{1,2},\ptmiss) < 0.4\\[-0.4ex]
              \text{no cut on} ~\met \end{array}$ & Control region A & Control region B \\[2.5ex]
        $\begin{array}{l}\Delta\phi(\text{jet}_{1,2},\ptmiss) > 0.4\\[-0.4ex]
              \met < \GeV{200/300}\end{array}$ & Control region C & Region D \\
    \hline
    \hline
  \end{tabular}}

  \subtable[Regions used for normalization and validation of the multijet pseudo-data
            in the \twotau analysis. The \met object in the selection is defined by
            the jet-smearing method.\smallskip]{
 \renewcommand\arraystretch{1.17}
 \newcommand\makebroader{\rule{5ex}{0pt}}
 \begin{tabular}{c|c}
  \hline
    \hline
  \makebroader Multijet CR\hspace*{3ex}	& \hspace*{3ex} Multijet VR\makebroader\\
    \hline
  \multicolumn{2}{c}{$\LJetPt >\unit[130]{GeV}$, $\SLJetPt>\unit[30]{GeV}$}\\[-0.5ex]
  \multicolumn{2}{c}{$\met>\unit[150]{GeV}$}\\[-0.5ex]
  \multicolumn{2}{c}{$N_\ell^\text{baseline}=0$}\\[-0.5ex]
 \multicolumn{2}{c}{$\Delta\phi(\text{jet}_{1,2},\ptmiss) < 0.3$}\\[-0.5ex]
 \multicolumn{2}{c}{$\met/\meff < 0.4$}	\\
    \hline
  \makebroader$N_\tau^\text{loose}=0$\quad &\quad$N_\tau^\text{loose}=1$\makebroader\\
    \hline
    \hline
    \end{tabular}
}
  
  \end{table}

\subsubsection*{\texorpdfstring{\twotau}{Two-tau} channel}

Background events from multijet production contain both fake \met
from instrumental effects in the jet energy measurements and fake taus.
Since both effects are difficult to simulate reliably and the large
cross section would require very large simulation samples, the multijet background expectation for the \twotau final state is computed using a sample from data with the ``Jet Smearing'' 
technique \cite{PhysRevD.87.012008}. Using this method a sample of events with artificial \met is obtained, where all other particles, including fake taus, are taken from data. This sample
is then used in the analysis to estimate the background from multijet events.
Events with low \met are selected from data requiring that they pass a single-jet trigger and have an \met significance 
$S=\met/\sqrt{\sum E_\mathrm{T}} < \unit[0.6]{GeV^\frac{1}{2}}$, where $\sum {E_\mathrm{T}}$
includes the same reconstructed objects used for computing \met, as detailed in \sect~\ref{sec:ObjectReco}. 
A pseudo-data sample with fake \met is then obtained by applying jet energy resolution smearing to all jets in these events.
After subtracting the small contribution ($<7\%$) from other backgrounds using scaled MC simulations, this sample is 
normalized in a multijet-enriched CR defined by the criteria in \tab~\ref{tab:QCDCRs}, which include the presence of two or 
more jets with the same \pt{} requirements as the SR.

The performance of the method is assessed in a validation region (VR)
which has identical kinematic requirements to the normalization region
but where one tau is required (\tab~\ref{tab:QCDCRs}). All relevant kinematic properties, including
those of the fake taus, are found to be well described by the normalized
multijet template, as shown in \fig~\ref{fig:valRegion2tau}(b)
for one of the kinematic variables considered in the analysis.

\subsubsection*{\texorpdfstring{\taulep}{Tau+lepton} channel}

In the \taulep channels the background contribution due to events with fake leptons is dominated by multijet events. Hence the multijet background contribution can be obtained from data
by estimating the number of fake lepton events. For this purpose, the ``matrix method'' described in \refer~\cite{ATLAS:SusyLep7TeV} is used, which exploits the difference 
in the isolation of the lepton candidates in events with true and fake leptons. The estimated contribution is found to be negligible.


\section{Systematic uncertainties on the background}
\label{sec:syst}
%
Various systematic uncertainties were studied and the effect
on the number of expected background events in each of the SRs
was calculated. 
Because of the normalization procedure in the CRs, these estimates are not
affected by theoretical errors on absolute cross sections, but only by generator
dependencies when extrapolating from the CRs to the SRs.

The difference in the estimated number of background events from two different 
generators is used to define the uncertainty due to the choice of MC generator
for the \ttbar, $W$+jets, $Z$+jets and diboson samples (see \sect~\ref{sec:SimulatedSamples}).
Moreover, the uncertainties on initial- and final-state radiation modelling and renormalization 
and factorization scales, which are found to be relatively small, are fully covered by the difference in generators.
For all samples, the statistical uncertainty on the prediction obtained 
from the alternative MC generator is also included in the estimate of the generator uncertainty. 

The experimental systematic uncertainties on the SM background estimates
arise from the jet energy scale and resolution~\cite{ATLAS:JES2014},
the tau energy scale~\cite{ATLAS:TES2013} and tau identification~\cite{ATLAS:TauID2013}.
The relative difference between the number of expected background events obtained with
the nominal MC simulation and that obtained after applying the uncertainty
variations on the corresponding objects is taken to be the systematic uncertainty on the background estimate. The uncertainties from the jet and tau energy scales are the largest
experimental uncertainties and are treated as uncorrelated, given that they are 
calibrated by different methods. 
The systematic uncertainty associated with the simulation of pile-up is taken into account by recomputing the event weights in all MC samples such that the resulting variation in the average interactions per bunch crossing corresponds to the observed uncertainty.
The uncertainty on the integrated luminosity is 2.8\%, as detailed in \refer~\cite{Aad:2013ucp}. 
This uncertainty affects only the normalization of the diboson background,
which is estimated entirely from simulation.

Additional uncertainties due to the methods used to estimate the background from multijet events are also considered.
In the \onetau channel, a 100\% uncertainty is obtained by taking into account possible correlations between the variables
used in the ABCD method, as well as the uncertainties on the scaling factors of the 
non-multijet samples that are subtracted from the data.
In the \twotau channel, uncertainties of the Jet Smearing method are evaluated by 
varying the jet response function used within the smearing process. This reflects 
the uncertainty on the ability to constrain the jet response to data in special 
multijet control regions when measuring the optimal jet response~\cite{PhysRevD.87.012008}.
In the \taulep channels, given that only an upper limit on the estimate
of the multijet background is obtained, a conservative 100\% uncertainty
on the multijet background is assumed.

\begin{table*}[!tp]
\begin{center}
\caption{Overview of the major systematic uncertainties on the total expected background in each signal region for the background 
estimates in the channels presented in this paper. The total systematic error also includes some minor systematic uncertainties, 
not detailed in the text or in the table.}
\label{tab:systematicErrors}
\medskip
\begin{tabular}{l|c c|c c c c c}
\hline
\hline
Source of uncertainty  & \onetau Loose & \onetau Tight & \twotau Incl.
                       & \twotau GMSB  & \twotau nGM   & \twotau bRPV \\
\hline
Generator uncertainties&  19\% &  30\% &  22\% &  78\% &  27\% &  33\% \\
Jet energy resolution  & 2.8\% & 9.7\% & 2.1\% & 4.7\% & 2.1\% & 9.4\% \\
Jet energy scale       & 3.6\% & 4.0\% & 5.3\% & 2.4\% & 4.9\% & 8.0\% \\
Tau energy scale       & 3.6\% & 1.3\% & 2.3\% & 8.6\% & 3.0\% & 2.8\% \\
Pile-up re-weighting    & 1.0\% & 1.0\% & 1.4\% & 1.5\% & 1.6\% & 1.3\% \\
Multijet estimate      &10.5\% & 9.6\% & 2.0\% & 7.5\% & 0.8\% & 3.8\% \\
\hline
Total syst.            & 24\%  & 35\%  & 24\%  & 79\%  & 30\%  & 36\% \\
\hline
\hline
\end{tabular}\bigskip%

\begin{tabular}{l|c c c c|c c c c}
\hline
\hline
Source of uncertainty  & \tauel & \tauel & \tauel & \tauel 
                       & \taumu & \taumu & \taumu & \taumu \\
                       & {\small GMSB}   & {\small nGM}    & {\small bRPV}   & {\small mSUG.} 
                       & {\small GMSB}   & {\small nGM}    & {\small bRPV}   & {\small mSUG.} \\
\hline
Generator uncertainties &51\% &46\% &19\% &28\% &28\% &30\% &39\% &32\% \\
Jet energy resolution  & 4\% & 5\% & 9\% & 3\% & 5\% & 6\% & 8\% & 3\% \\
Jet energy scale       & 7\% & 9\% & 7\% &12\% & 7\% &13\% &10\% &13\% \\
Tau energy scale       & 7\% & 2\% & 8\% & 1\% & 8\% & 8\% & 4\% & 4\% \\
Pile-up re-weighting    & 3\% & 2\% & 1\% & 0\% & 2\% & 3\% & 1\% & 1\% \\
\hline
Total syst.            &60\% &48\% &32\% &30\% &36\% &34\% &41\% &33\% \\
\hline
\hline
\end{tabular}
\end{center}
\end{table*}%
The total systematic uncertainty related to the background estimation
and its breakdown into the main contributions are shown in \tab~\ref{tab:systematicErrors}
for each signal region.

The total experimental systematic uncertainty on the signal selection efficiency from the various sources discussed in this section varies for each channel and for each signal
model considered. In the GMSB scenario this uncertainty is 5--10\% for the \onetau channel, rising to 20\% for high values of $\Lambda$; 
20--30\% for most of the parameter space in the \twotau channel, increasing to as high as 45\% in the region of highest $\Lambda$ and low $\tan\beta$;
5--15\% for the \taulep channel.
In the mSUGRA model the signal systematic uncertainty is at the level of 10\% across most of the $(\mz,\moh)$ plane for all channels.
The total experimental uncertainty on the signal selection efficiency in the nGM scenario is
10--20\% for the \twotau channel; in the \taulep channels it is of the order of 15--20\% for lower masses
and decreases to an average level of 5--10\% for high $m_{\tilde{g}}$.
In the $(\mz,\moh)$ plane of the bRPV model the total systematic uncertainty on the signal selection efficiency
is at the level of 10\% across most of the plane for all channels, rising to 50\% at the lowest $m_{1/2}$ region studied 
and to 80\% for individual signal samples generated at the highest $m_{1/2}$ values.

\section{Results}
\label{sec:Results}
\subsection*{Observed data and expected background events in the signal regions}
\begin{figure}[!tp]
  \begin{center}
   \subfigure[\twotau \mTT distribution]{\hspace*{-1ex}\includegraphics[width=0.51\textwidth]{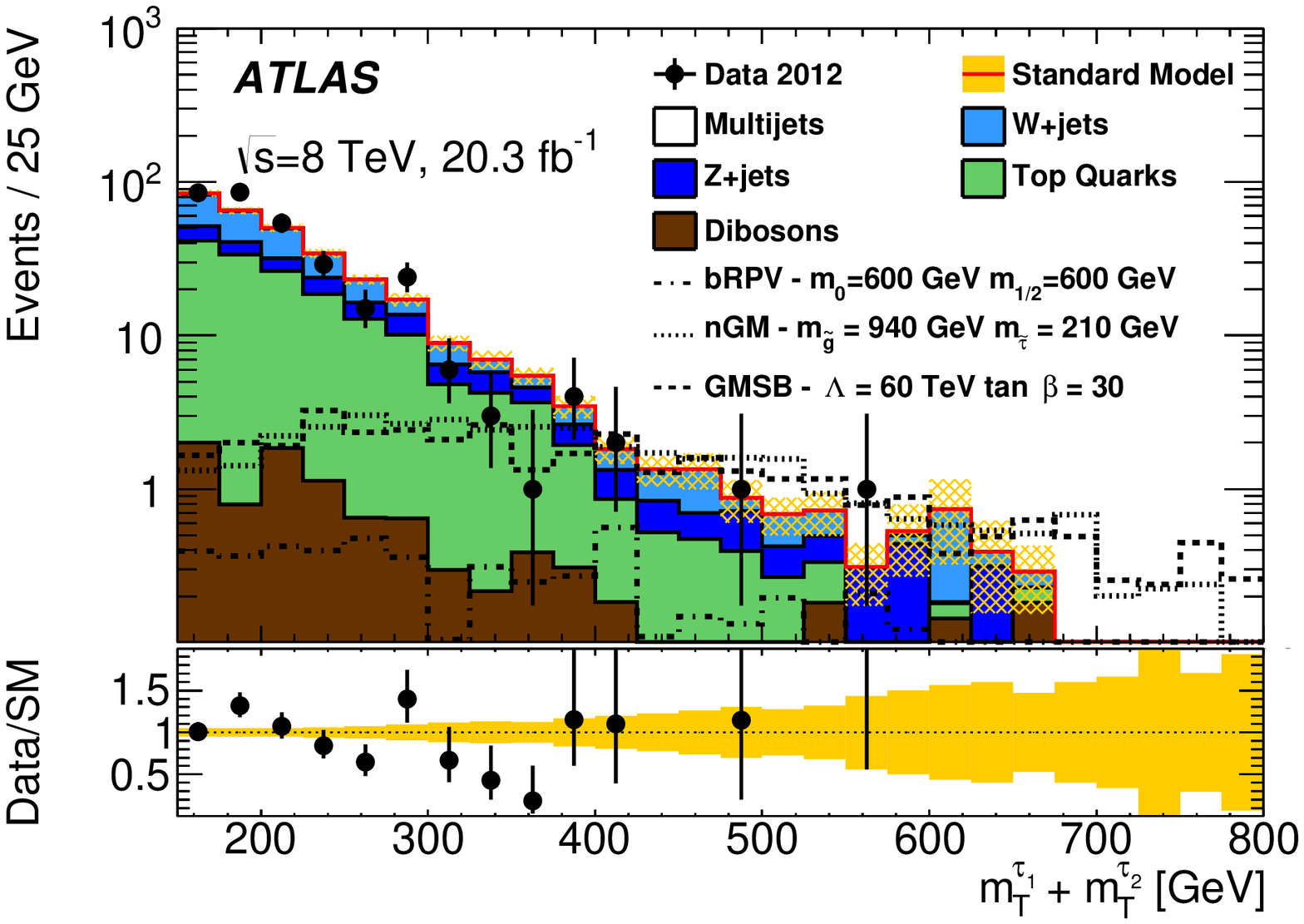}}
   \subfigure[\twotau \HTtj distribution]{\includegraphics[width=0.51\textwidth]{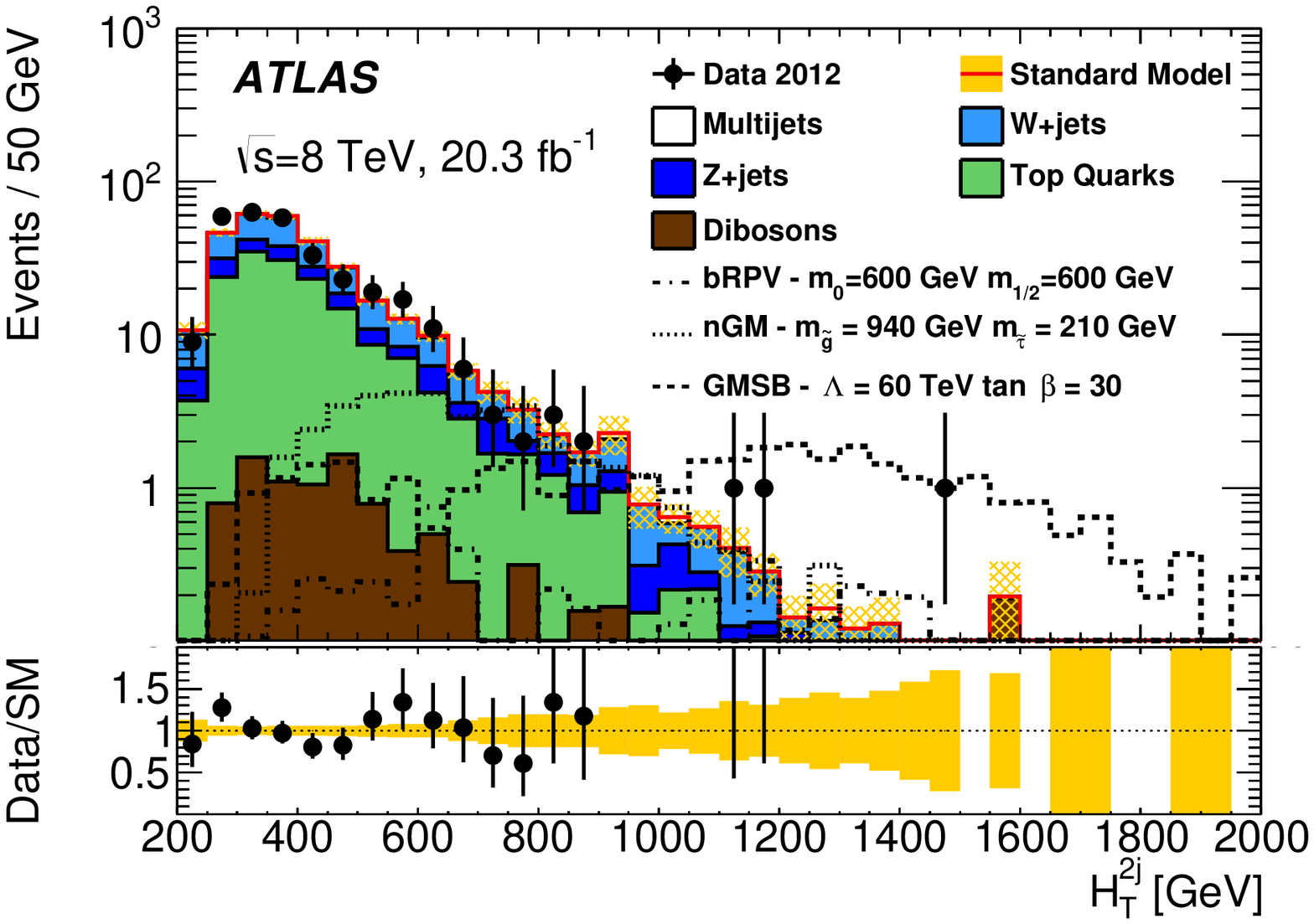}\hspace*{-3ex}}
   \subfigure[\twotau \njet distribution]{\vspace*{-1ex}\includegraphics[width=0.51\textwidth]{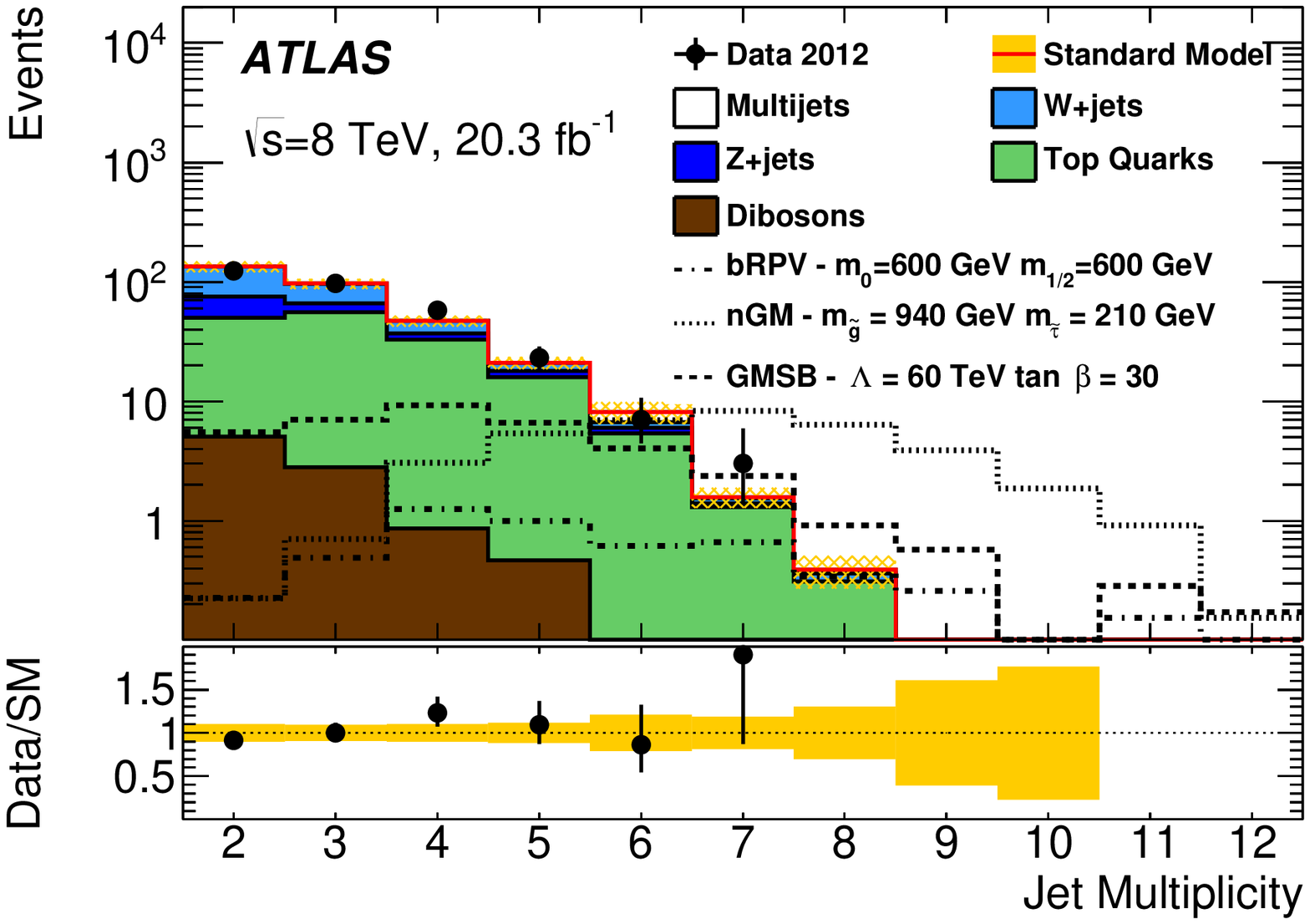}}
    \end{center}
  \caption{Distribution of \mTT{}, \HTtj{} and \njet in the \twotau channel after all analysis requirements but the final SR requirements on \mTT{} and \HTtj{}.
    To reduce the contributions from events with $Z$ bosons decaying into tau leptons, the requirement $\mTT>\unit[150]{GeV}$ is applied to all distributions.
    Data are represented by the points. The SM prediction includes the data-driven corrections discussed in the text.
    The shaded band centred around the total SM background indicates the statistical uncertainty on the background expectation.
    MC events are normalized to data in the CRs corresponding to \HTtj{} below \unit[550]{GeV}.
    Also shown is the expected signal from typical bRPV, nGM and GMSB samples.
    There are no data events in the overflow bin after all analysis requirements are applied
  }
  \label{fig:2tauResults}
\end{figure}

\begin{figure}[h!]
  \begin{center}
    \subfigure[\tauel bRPV SR, $\njet\geq4$]{\hspace*{-2ex}\includegraphics[width=0.51\textwidth,bb=0 0 567 384]{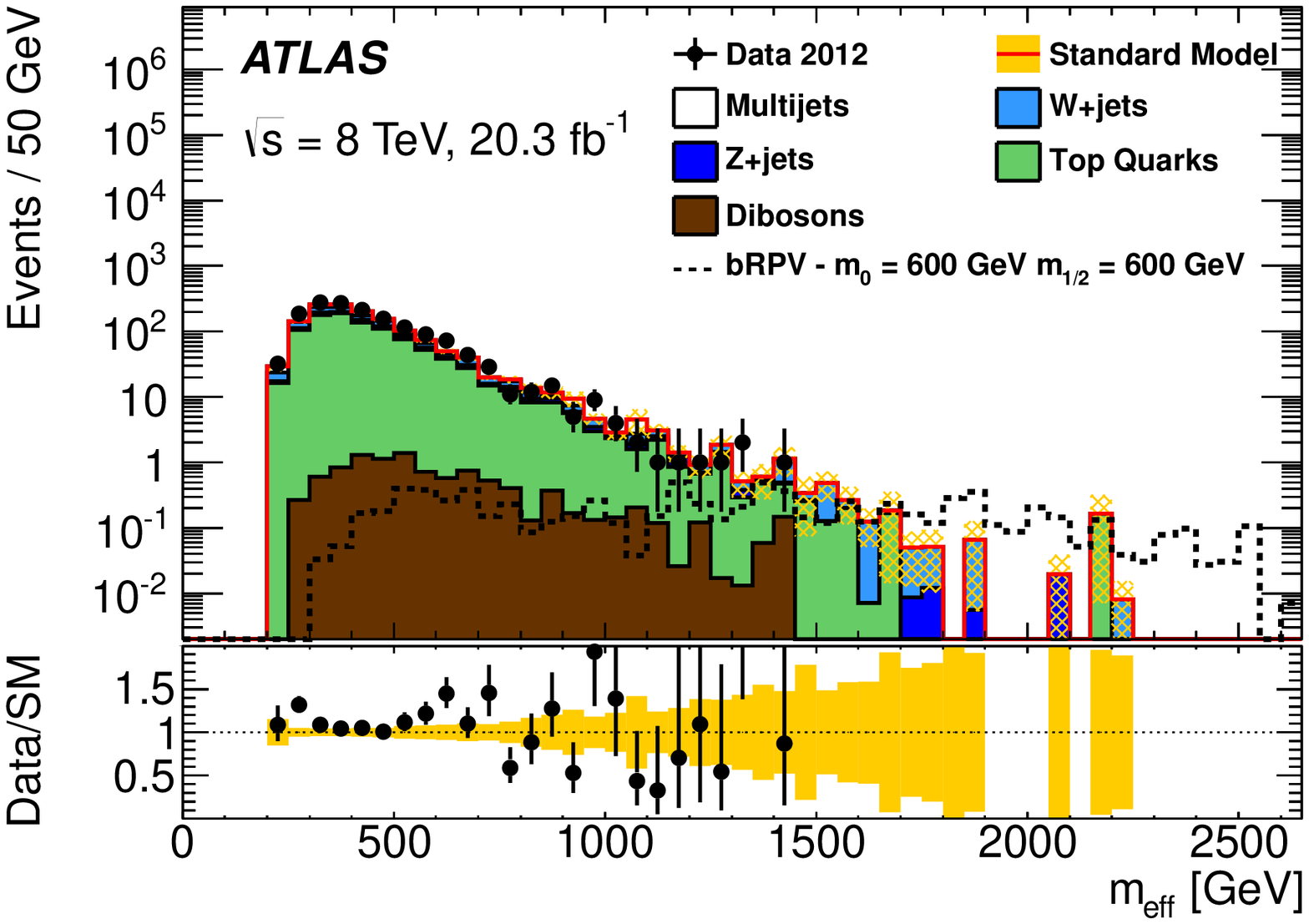}}
    \subfigure[\tauel GMSB SR]{\includegraphics[width=0.51\textwidth,bb=0 0 567 384]{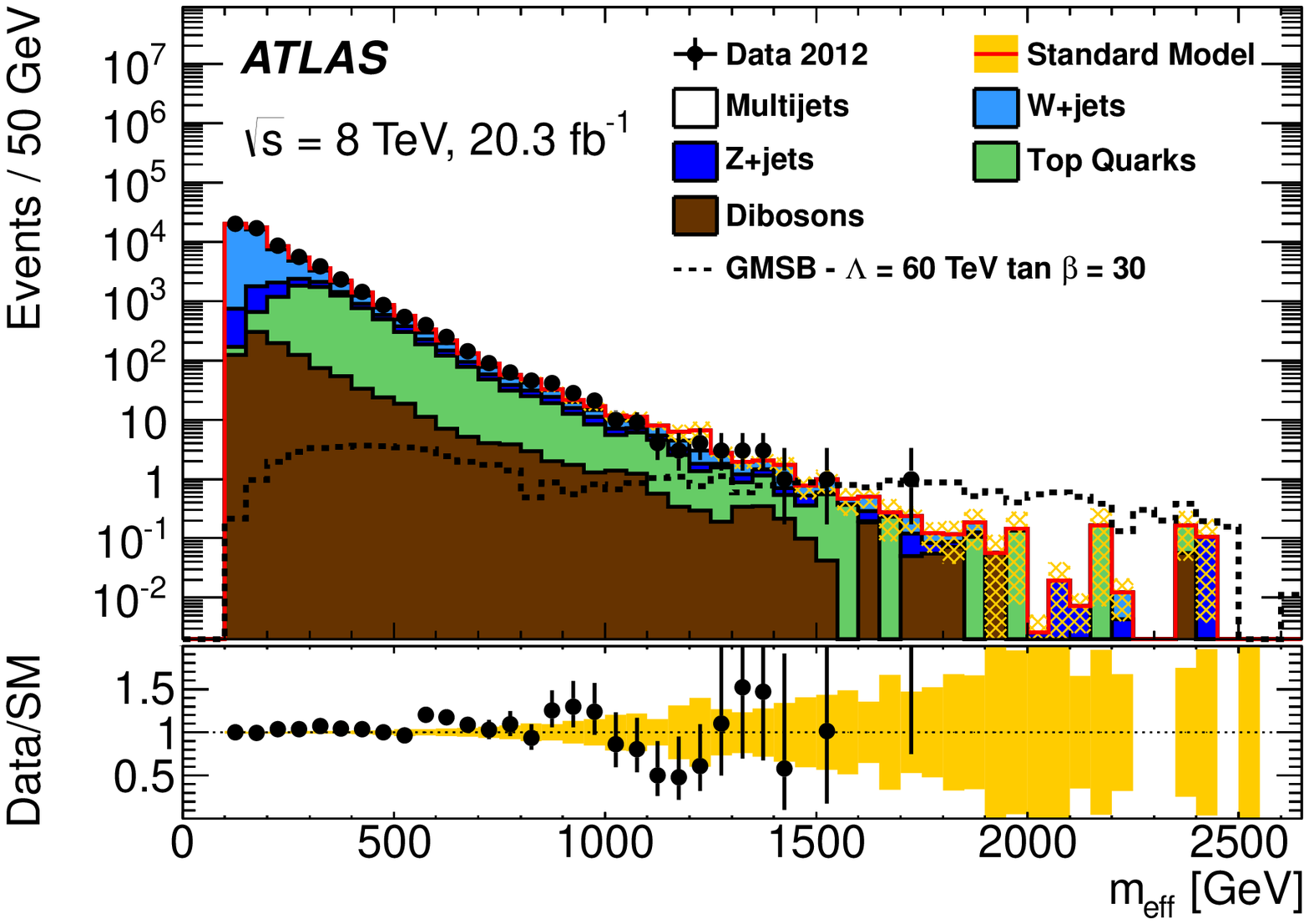}\hspace*{-2ex}}
    \subfigure[\tauel mSUGRA SR, $\njet\geq3$]{\hspace*{-2ex}\includegraphics[width=0.51\textwidth,bb=0 0 567 384]{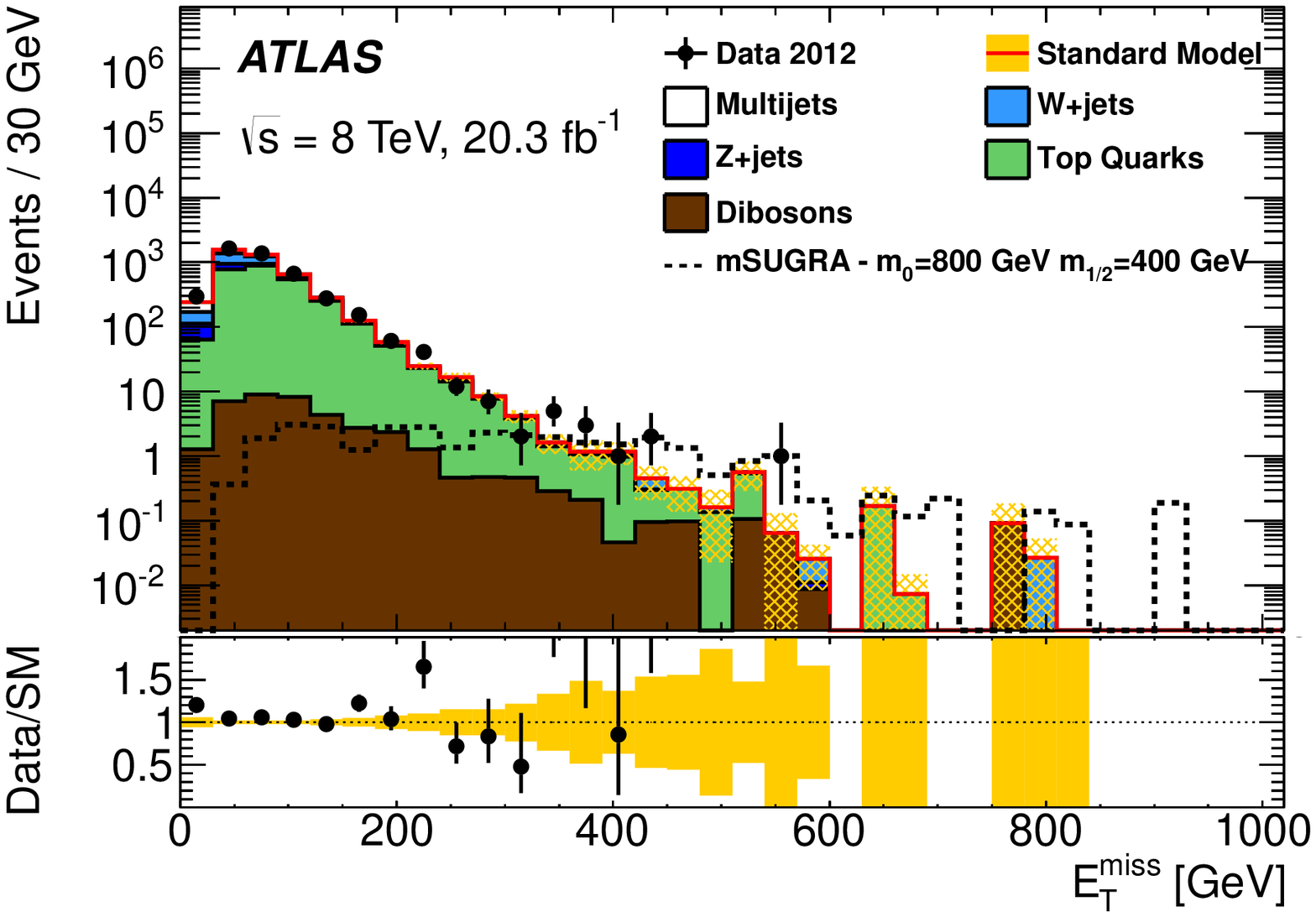}}
    \subfigure[\tauel nGM SR, $\njet\geq3$]{\includegraphics[width=0.51\textwidth,bb=0 0 567 384]{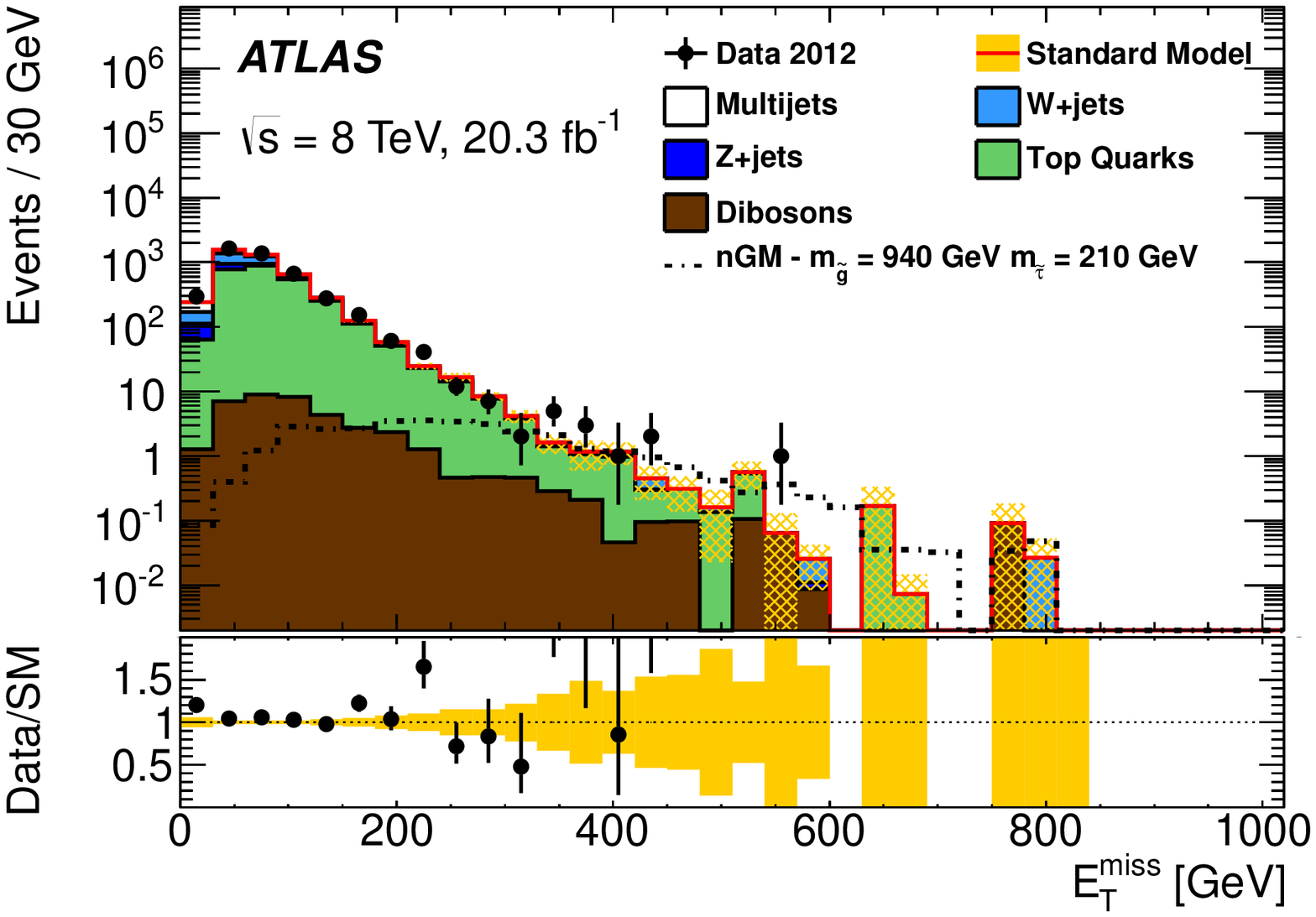}\hspace*{-2ex}}
  \end{center}
  \caption{Distribution of the final kinematic variables in the \tauel channel after all analysis requirements but the final SR selections on \meff and \met.
    Data are represented by the points. The SM prediction includes the data-driven corrections discussed in the text.
    The shaded band centred around the total SM background indicates the statistical 
    uncertainty on the background expectation.
    MC events are normalized to data in the CRs described in the text.
    Also shown is the expected signal from typical bRPV, GMSB, mSUGRA and nGM signal samples.
    The last bin in the expected background distribution is an overflow bin.
    There are no data events in the overflow bin after all analysis requirements are applied.}
    \label{fig:tauelResults}
\end{figure}

\begin{figure}[h!]
  \begin{center}
    \subfigure[\taumu bRPV SR, $\njet\geq4$]{\hspace*{-2ex}\includegraphics[width=0.51\textwidth,bb=0 0 567 384]{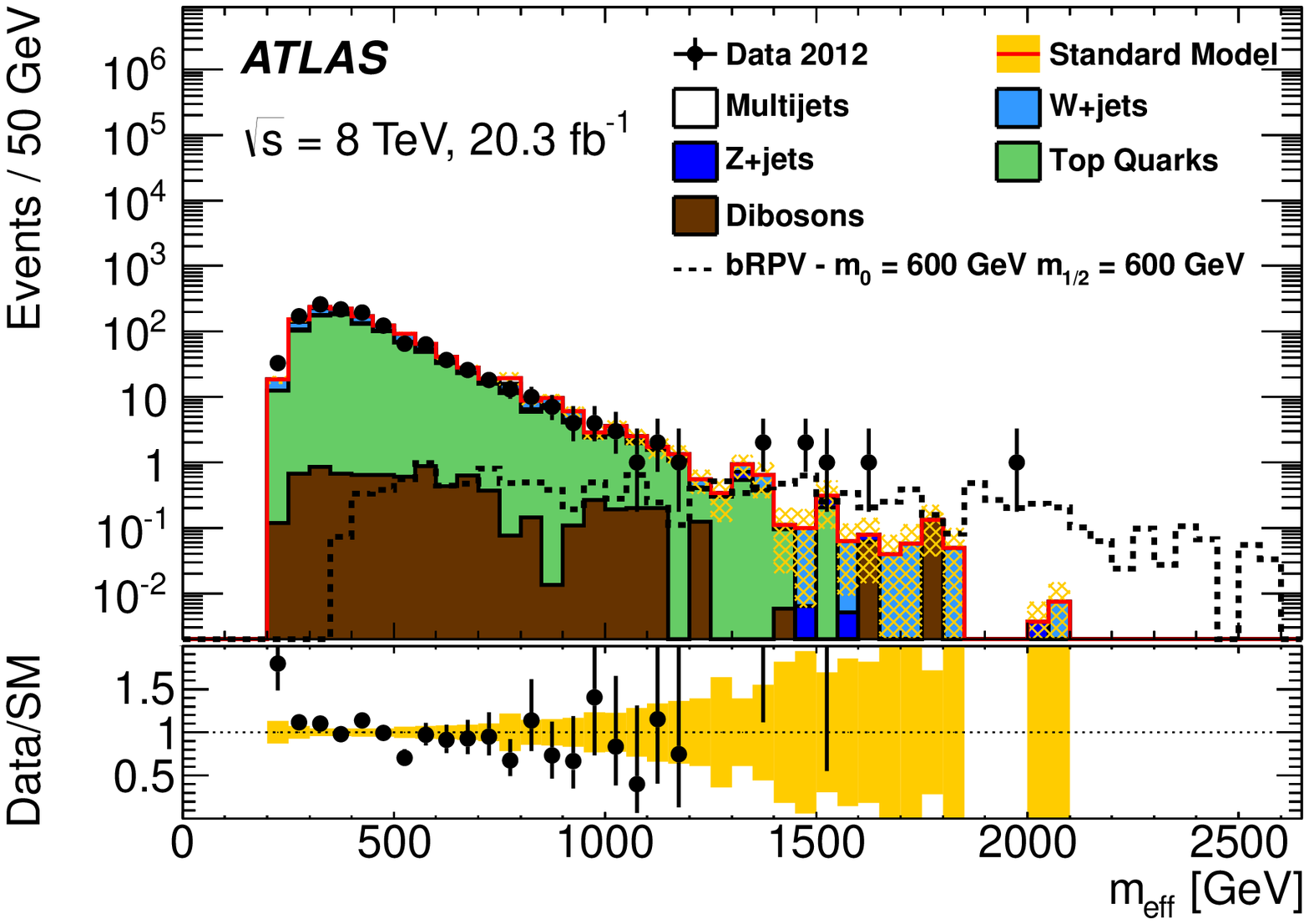}}
    \subfigure[\taumu GMSB SR]{\includegraphics[width=0.51\textwidth,bb=0 0 567 384]{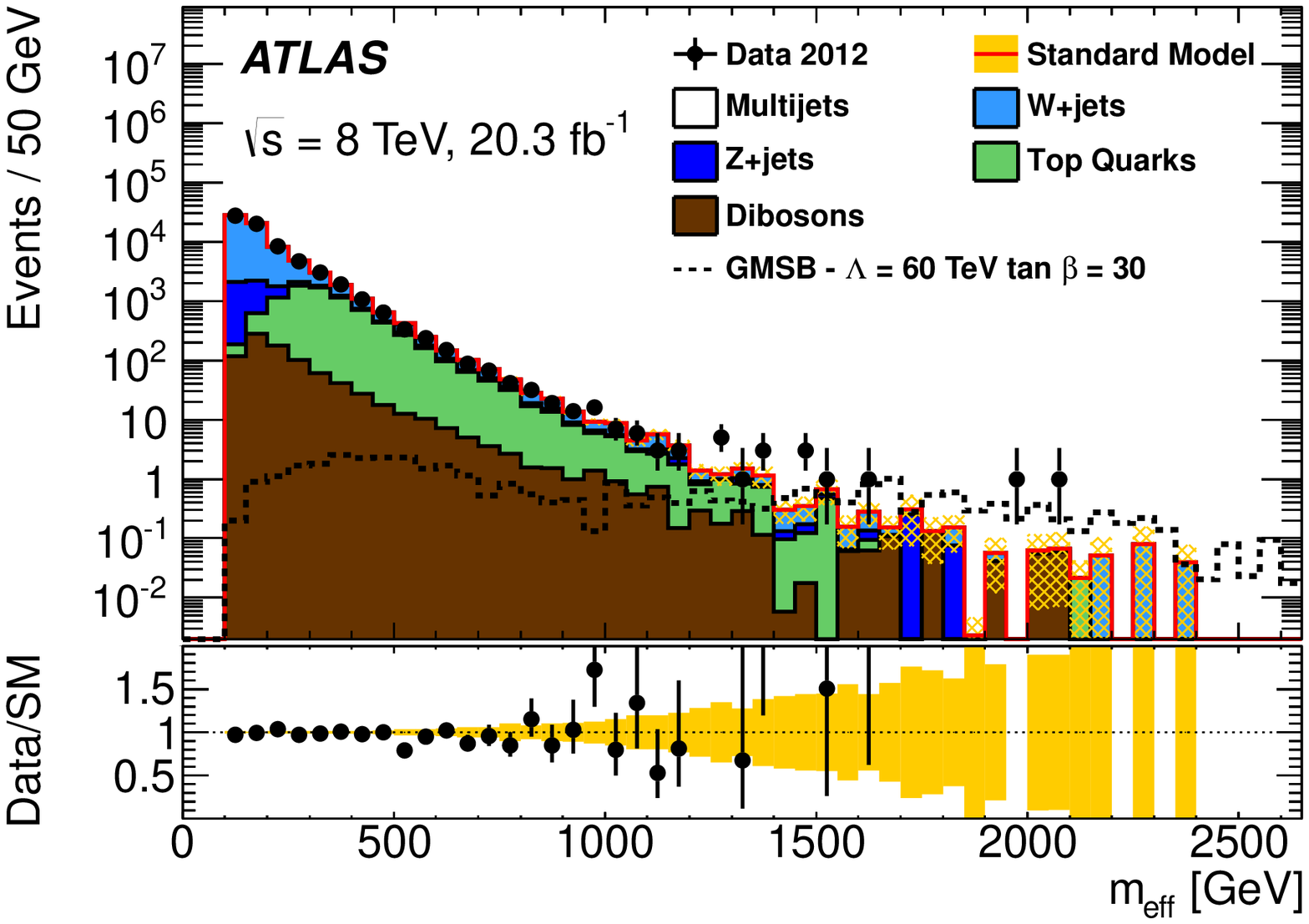}\hspace*{-2ex}}
    \subfigure[\taumu mSUGRA SR, $\njet\geq3$]{\hspace*{-2ex}\includegraphics[width=0.51\textwidth,bb=0 0 567 384]{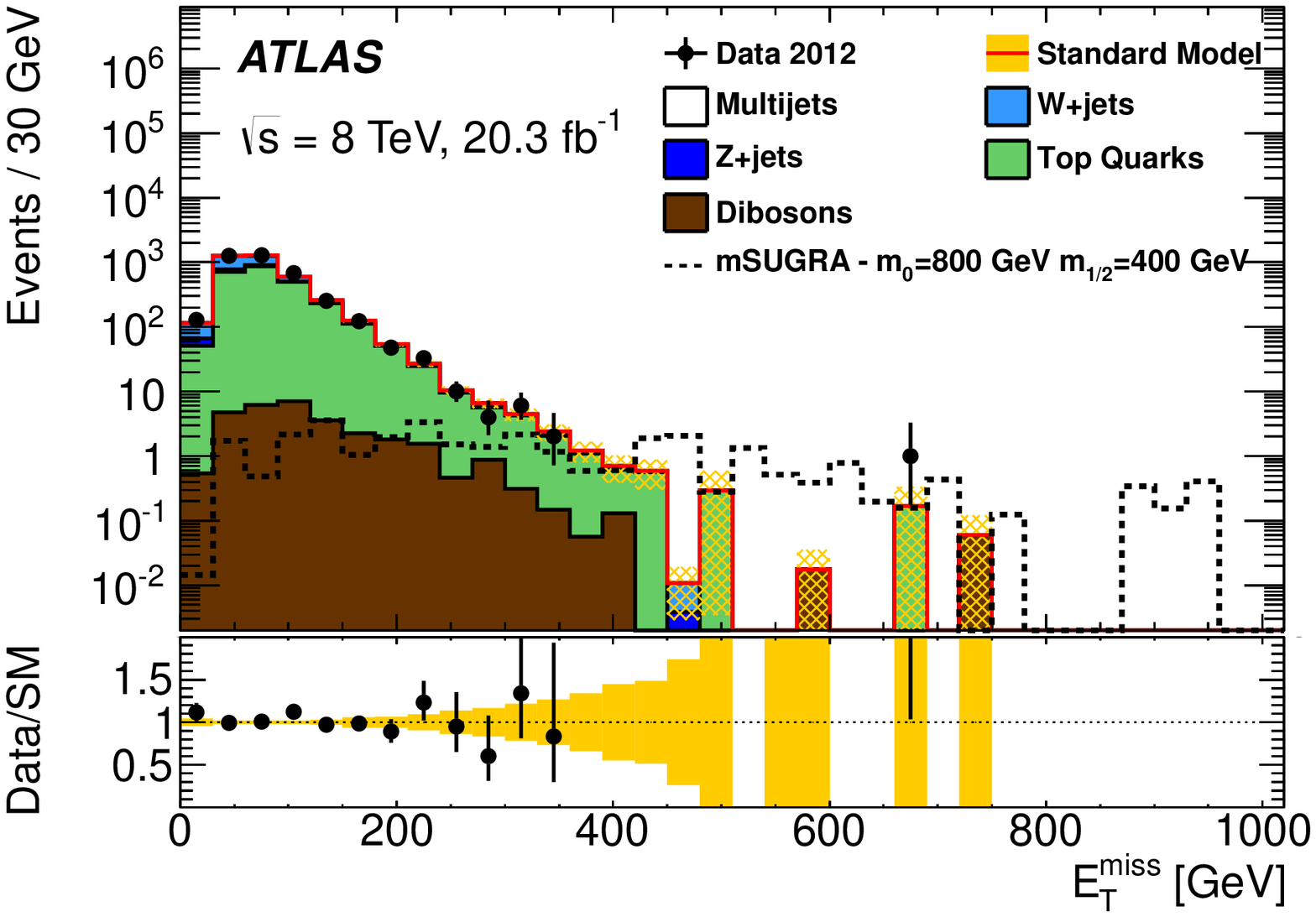}}
    \subfigure[\taumu nGM SR, $\njet\geq3$]{\includegraphics[width=0.51\textwidth,bb=0 0 567 384]{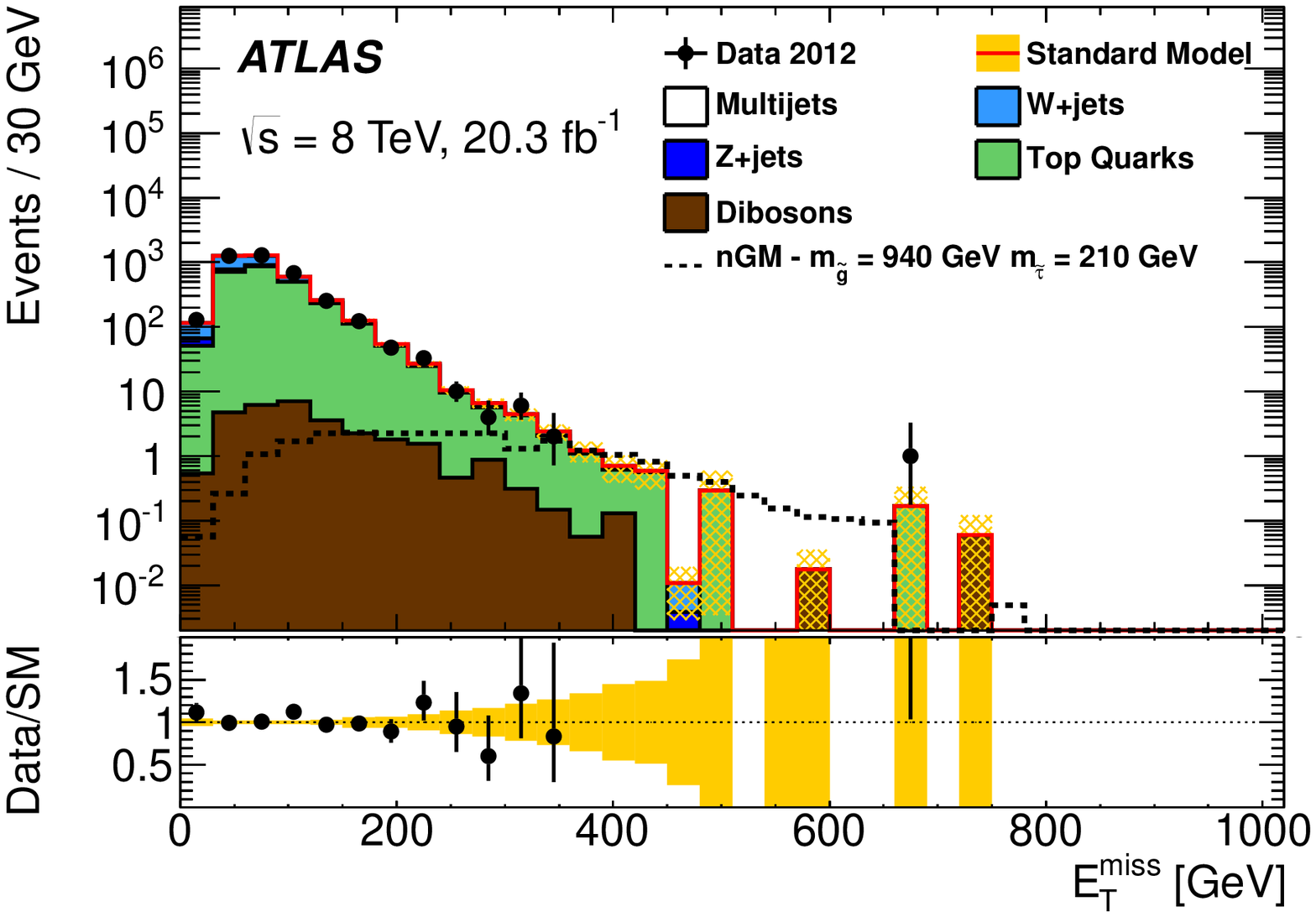}\hspace*{-2ex}}
  \end{center}
  \caption{Distribution of the final kinematic variables in the \taumu channel after all analysis requirements but the final SR selections on \meff and \met.
    Data are represented by the points. The SM prediction includes the data-driven corrections discussed in the text.
    The shaded band centred around the total SM background indicates the statistical 
    uncertainty on the background expectation.
    MC events are normalized to data in the CRs described in the text.
    Also shown is the expected signal from typical bRPV, GMSB, mSUGRA and nGM signal samples.
    The last bin in the expected background distribution is an overflow bin.
    There are no data events in the overflow bin after all analysis requirements are applied.}
  \label{fig:taumuResults}
\end{figure}

Data and scaled background simulation were compared for different kinematic quantities.
\begin{figure}[h!]
  \begin{center}
   \subfigure[\onetau Loose SR, $\met>\GeV{200}$]{\hspace*{-2ex}\includegraphics[width=0.51\textwidth,bb= 0 0 600 400]{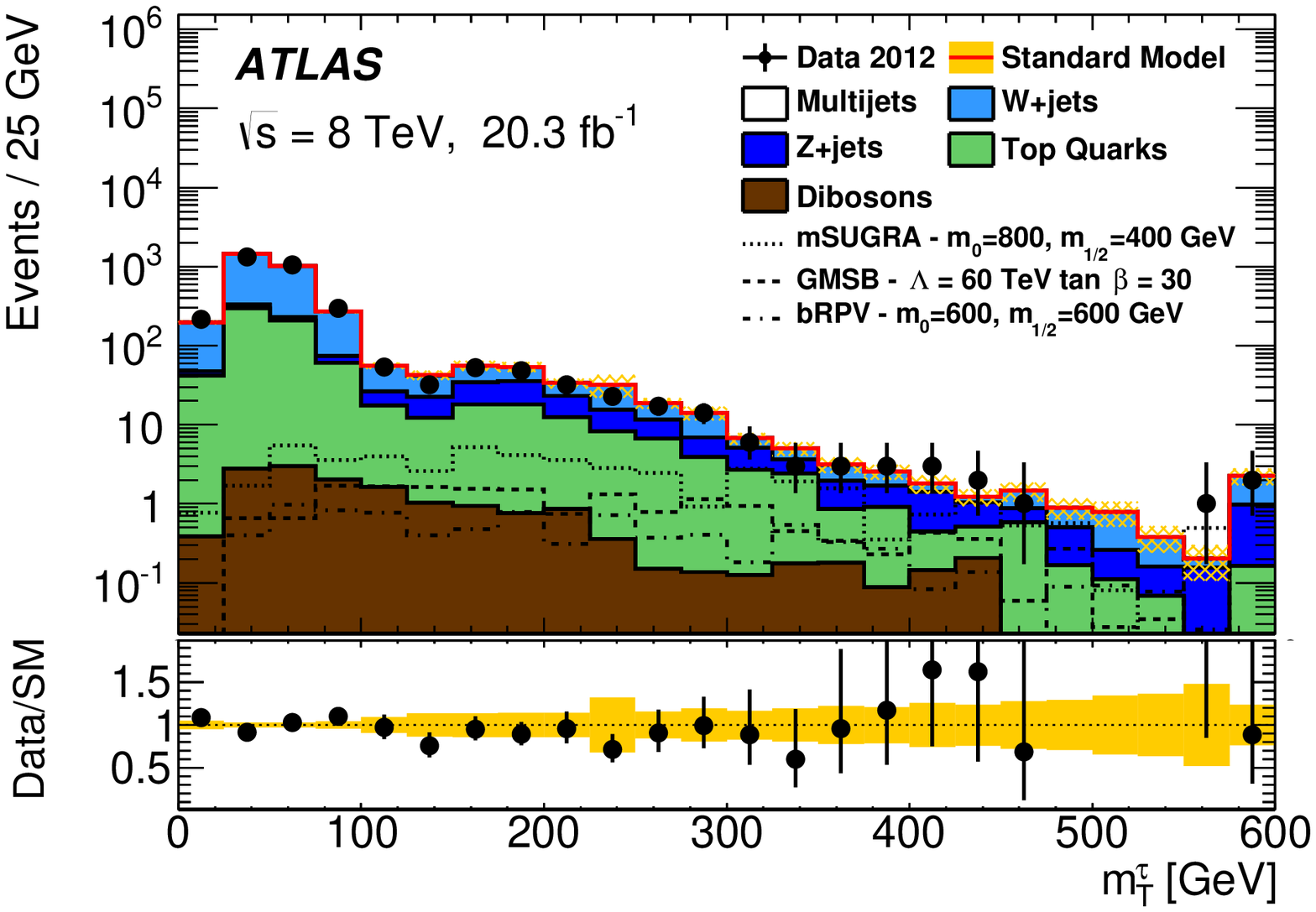}}
   \subfigure[\onetau Loose SR, $\met>\GeV{200}$]{\includegraphics[width=0.51\textwidth,bb=0 0 600 400]{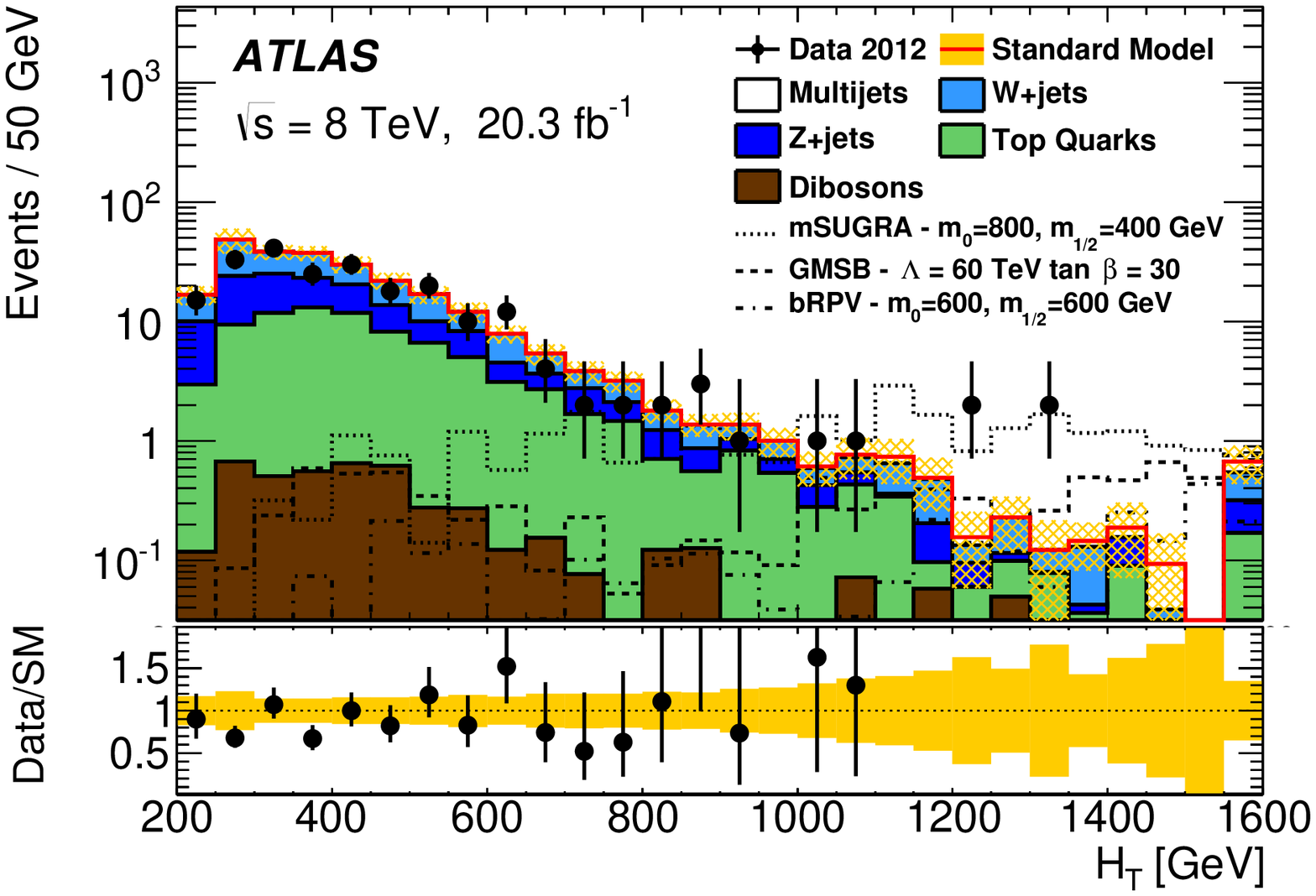}\hspace*{-2ex}}
   \subfigure[\onetau Tight SR, $\met>\GeV{300}$]{\hspace*{-2ex}\includegraphics[width=0.51\textwidth,bb=0 0 600 400]{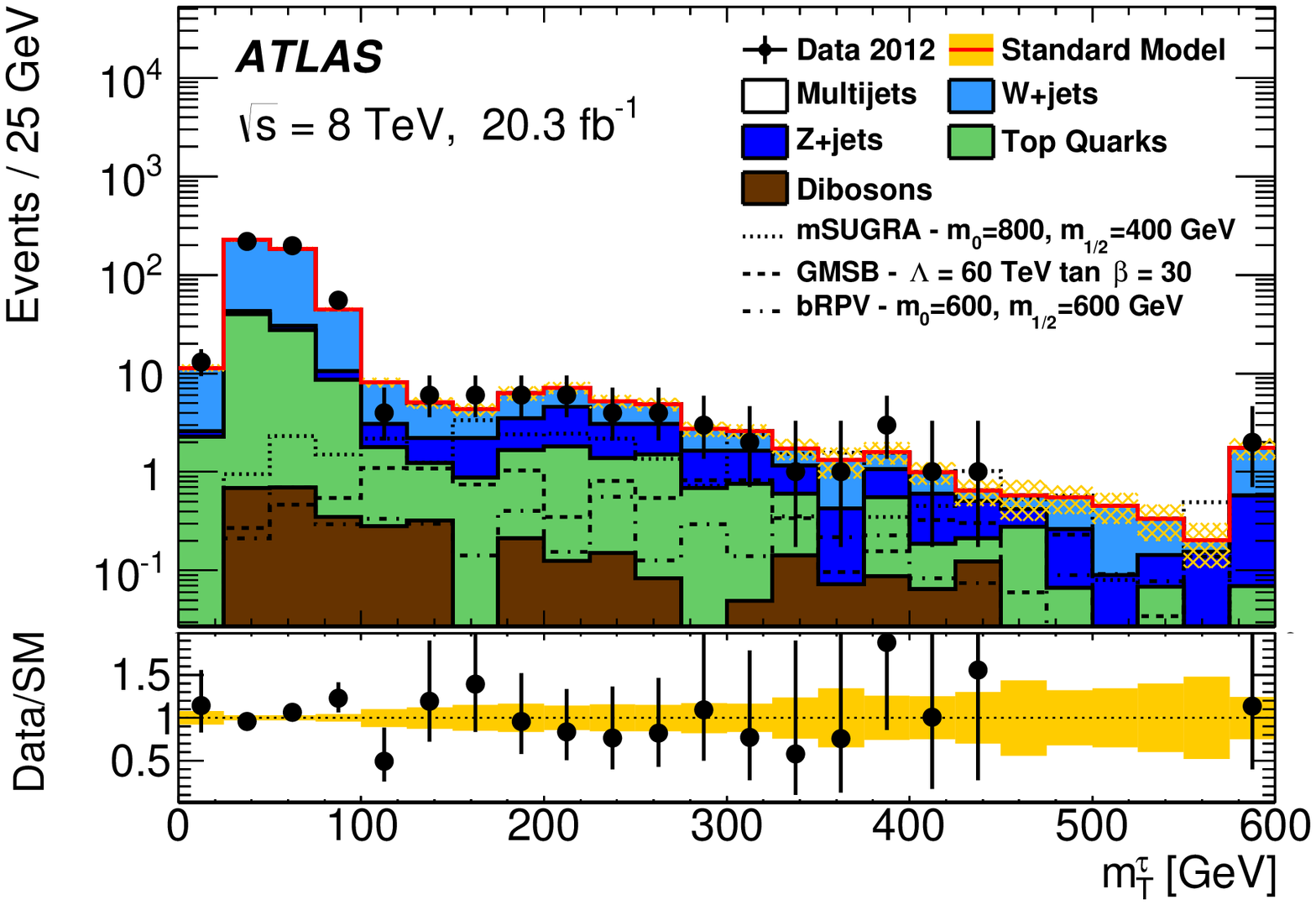}}
   \subfigure[\onetau Tight SR, $\met>\GeV{300}$]{\includegraphics[width=0.51\textwidth,bb=0 0 600 400]{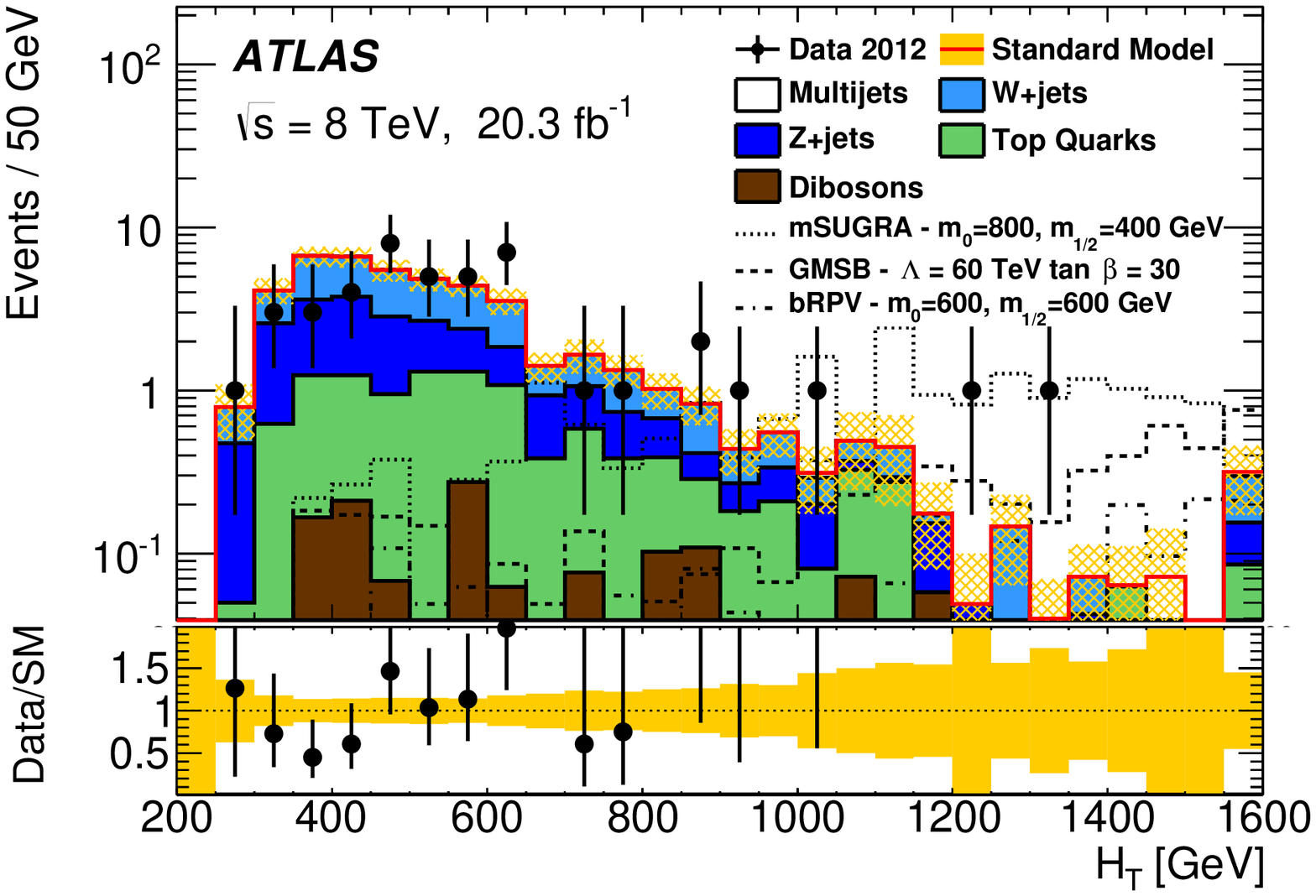}\hspace*{-2ex}}
  \end{center}
  \caption{Distribution of \mT{} after all analysis requirements but the requirement on \mT{} and the final requirement on \HT{}, and of \HT{} after 
    the \mT{} requirement for (a,\,b) the \onetau ``Loose'' and (c,\,d) ``Tight'' SRs. 
    Data are represented by the points. The SM prediction includes the data-driven corrections discussed in the text.
    The shaded band centred around the total SM background indicates the statistical uncertainty on the background expectation.
    MC events are normalized to data in the CRs corresponding to \mT{} below \unit[130]{GeV}. 
    Also shown is the expected signal from typical mSUGRA, GMSB and bRPV samples.
    The last bin in the expected background distribution is an overflow bin.
  }
  \label{fig:1tauResults}
\end{figure}%
\Fig~\ref{fig:1tauResults} shows the \mT{} distribution after all the requirements of the analysis except the ones on \mT{} and \HT{}, as well as 
the \HT{} distribution after the requirement on \mT{} for the \onetau channel. ``Loose'' and ``Tight'' SR plots are displayed individually with the corresponding 
requirement on \met applied.
\Fig~\ref{fig:2tauResults} shows the \mTT{}, \HTtj{} and \njet distributions after all the requirements of the analysis except the final selection on \mTT{} and \HTtj{} 
for the \twotau channel. The $\mTT>\unit[150]{GeV}$ requirement common 
to all SRs is applied to reduce contributions from events with $Z$ bosons decaying into tau leptons.
\Figs~\ref{fig:tauelResults} and \ref{fig:taumuResults} show the \meff and \met\ distributions for each of the SRs in the \taulep channels. All common requirements and the jet 
multiplicity selection corresponding to the respective SR are applied.

Good agreement between data and SM expectations is observed for all distributions after applying all corrections and data-driven background estimation techniques. 

\Tabs~\ref{tab:eventYields_onetau}--\ref{tab:eventYields_taumu} summarize the number of observed events in the four channels in data and the number of expected background events.
No significant excess over the Standard Model background estimate is observed.
Upper limits at 95\%~confidence level (CL) on the number of signal events for each SR independent of any specific SUSY model are derived using the CLs prescription~\cite{Read:2002hq}.
The profile likelihood ratio is used as a test statistic~\cite{asimov} and all systematic uncertainties on the background estimate are treated as nuisance parameters,
neglecting any possible signal contamination in the control regions. 
The limits are computed by randomly generating a large number
of pseudo-datasets and repeating the CLs procedure. This calculation was compared
to an asymptotic approximation~\cite{asimov}, which is used for
the model-dependent limits, and was found to be in good agreement.
These limits are then translated into upper limits on the visible signal cross section,
$\sigma_\text{vis}$, by normalizing them to the total integrated luminosity in data. 
The visible cross section is defined as 
the product of acceptance, selection efficiency and production cross section. These results are also given in \tabs~\ref{tab:eventYields_onetau}--\ref{tab:eventYields_taumu} for all channels.

\begin{table*}[!htb]
\begin{center}
    \caption{Number of expected background events and data yields in the \onetau final state.
             Where possible, the uncertainties on the number of expected events are separated into statistical (first) and systematic (second) components.
             The statistical uncertainty comprises the limited number of simulated events in both the SR and the CRs as well as the limited number of data events in the CRs.
             The SM prediction is computed taking into account correlations between the different uncertainties.
             Also shown are the number of expected signal events for one selected benchmark point for each signal model studied. 
             For GMSB the chosen point has the parameters \GMSBBenchPars, for nGM \nGMBenchPars, for bRPV \bRPVBenchPars and for mSUGRA \mSugraBenchPars.
	     \newline
             The resulting 95\%~\CL limit on the number of observed (expected) signal events and on the visible cross sections
             from any new-physics scenario for each of the final states is shown, taking into account the
             observed events in data and the background expectations. Discovery p-values are capped at $0.5$ in cases where the expected number of events exceeds the observed number.
\label{tab:eventYields_onetau}}\medskip
\footnotesize
\begin{tabular}{l|c|c}
\hline
\hline
 --			& \onetau Loose 		& \onetau Tight 		\\
\hline\hline
Multijet 		& \onetauLooseQcdEstim		& \onetauTightQcdEstim		\\
$W$ + jets		& \onetauLooseWEstim		& \onetauTightWEstim		\\ 
$Z$ + jets		& \onetauLooseZEstim		& \onetauTightZEstim		\\ 
Top			& \onetauLooseTopEstim		& \onetauTightTopEstim		\\ 
Diboson			& \onetauLooseDBEstim		& \onetauTightDBEstim		\\
\hline\hline
Total background 	& \onetauLooseTotalEstim 	& \onetauTightTotalEstim	\\
\hline\hline
Data       		& \onetauLooseData		& \onetauTightData		\\
\hline
\hline
Signal MC Events &\\
~~GMSB 60/30		& --				& \onetauTightGMSB		\\
~~nGM 940/210 		& -- 				& --				\\
~~bRPV 600/600		& --				& \onetauTightbRPV		\\
~~mSUGRA 800/400 	& --				& \onetauTightmSUGRA		\\
\hline
\hline
Obs (exp) limit &\\
on signal events 	& 11.7 (${10.1}^{+3.6}_{-2.6}$) & 5.9 ($5.3^{+1.8}_{-1.3}$)	\\
\hline
Obs (exp) limit on &\\
vis. cross section (fb)	& 0.58 (0.50)			& 0.29 (0.26)			\\
\hline
Discovery p-value&\\
$p(s=0)$		& 0.37				& 0.37 				\\
\hline
\hline
\end{tabular}

\end{center}
\end{table*}

\begin{table*}
\begin{center}
    \caption{Number of expected background events and data yields in the \twotau final state.
	      Further details can be found in \tab~\ref{tab:eventYields_onetau}.
\label{tab:eventYields_twotau}}\medskip
\footnotesize

\begin{tabular}{l|c|c|c|c}
\hline
\hline
 --			& \twotau Inclusive			& \twotau GMSB 			& \twotau nGM			& \twotau bRPV 			\\
\hline\hline
Multijet 		& \ditauQCDInclusive			& \ditauQCDGMSB			& \ditauQCDnGM			& \ditauQCDbRPVmoh			\\
$W$ + jets		& \ditauInclusiveW			& \ditauGMSBW				& \ditaunGMW			& \ditaubRPVmohW			\\ 
$Z$ + jets		& \ditauInclusiveZ			& \ditauGMSBZ				& \ditaunGMZ			& \ditaubRPVmohZ			\\ 
Top			& \ditauInclusiveTop			& \ditauGMSBTop			& \ditaunGMTop			& \ditaubRPVmohTop			\\ 
Diboson			& \ditauInclusiveDB			& \ditauGMSBDB				& \ditaunGMDB			& \ditaubRPVmohDB			\\
\hline\hline
Total background 	& \ditauInclusiveScaleStatSystNoLabel	& \ditauGMSBScaleStatSystNoLabel	& \ditaunGMScaleStatSystNoLabel	& \ditaubRPVmohScaleStatSystNoLabel	\\
\hline\hline
Data       		& \ditauInclusiveData			& \ditauGMSBData        		& \ditaunGMData		& \ditaubRPVmohData        		\\
\hline
\hline
Signal MC Events &&&\\
~~GMSB 60/30		& -- 					& \twotauGMSBBench			& -- 				& --					\\
~~nGM 940/210 		& -- 					& --					& \twotaunGMBench		& --					\\
~~bRPV 600/600		& -- 					& --					& -- 				& \twotaubRPVBench			\\
~~mSUGRA 800/400 	& -- 					& --					& -- 				& --					\\
\hline
\hline
Obs (exp) limit &&&\\
on signal events 	& 5.7 (${5.4}^{+1.7}_{-1.4}$) 	& 3.4 (${3.4}^{+0.6}_{-0.2}$)		& 3.8 (${5.4}^{+1.8}_{-1.5}$)	& 4.1 (${4.0}^{+1.5}_{-0.3}$)		\\
\hline
Obs (exp) limit on &&&\\
vis. cross section (fb) 	& 0.28 (0.26)& 0.17 (0.17)& 0.18 (0.27)& 0.20 (0.20)\\
\hline
Discovery p-value&&&\\
$p(s=0)$		& 0.47					& 0.50					& 0.50   			& 0.50				\\
\hline
\hline
\end{tabular}

\end{center}
\end{table*}

\begin{table*}[!hbp]
\begin{center}
    \caption{Number of expected background events and data yields in the \tauel final state.
	      Further details can be found in \tab~\ref{tab:eventYields_onetau}.
\label{tab:eventYields_tauel}}\medskip
\footnotesize

\begin{tabular}{l|c|c|c|c}
\hline
\hline
 --			& \tauel GMSB			& \tauel nGM 				& \tauel bRPV				& \tauel mSUGRA			\\
\hline\hline
Multijet 		& \tauelGMSBQCD		& \tauelNGMQCD				& \tauelBRPVQCD			& \tauelMSugQCD			\\
$W$ + jets		& \tauelGMSBW			& \tauelNGMW				& \tauelBRPVW				& \tauelMSugW				\\ 
$Z$ + jets		& \tauelGMSBZ			& \tauelNGMZ				& \tauelBRPVZ				& \tauelMSugZ				\\ 
Top		& \tauelGMSBTop		& \tauelNGMTop
& \tauelBRPVTop		& \tauelMSugTop			
\\ 
Diboson			& \tauelGMSBDiBosons		& \tauelNGMDiBoson			& \tauelBRPVDiBosons			& \tauelMSugDiBosons			\\
\hline\hline
Total background 	& \tauelGMSBAll		& \tauelNGMAll	& \tauelBRPVAll	& \tauelMSugAll	\\
\hline\hline
Data       		& \tauelGMSBData		& \tauelNGMData        		& \tauelBRPVData			& \tauelMSugData        		\\
\hline
\hline
Signal MC Events &&&\\
~~GMSB 60/30		& \tauelGMSBBench 				& --					& -- 					& --					\\
~~nGM 940/210 		& -- 				& \tauelNGMBench					& -- 					& --					\\
~~bRPV 600/600 		& -- 				& --					& \tauelBRPVBench 					& --					\\
~~mSUGRA 800/400	& -- 				& --					& -- 					& \tauelMSUGRABench					\\
\hline
\hline
Obs (exp) limit &&&\\
on signal events 	& 4.1 (${4.2}^{+1.7}_{-0.4}$) & 11.4 (${8.3}^{+2.8}_{-2.0}$)	& 5.3 (${6.0}^{+2.2}_{-1.1}$)  	& 14.6 (${11.7}^{+4.1}_{-3.2}$)	\\
\hline
Obs (exp) limit on &&&\\
vis. cross section (fb) 	& 0.20 (0.21)			& 0.56 (0.41)			& 0.26 (0.30)			& 0.72 (0.58)					\\
\hline
Discovery p-value&&&\\
$p(s=0)$		& 0.50				& 0.15					& 0.50   				& 0.24				\\
\hline
\hline
\end{tabular}

\end{center}
\end{table*}

\begin{table*}
\begin{center}
    \caption{Number of expected background events and data yields in the \taumu final state.
	      Further details can be found in \tab~\ref{tab:eventYields_onetau}.
\label{tab:eventYields_taumu}}\medskip
\footnotesize

\begin{tabular}{l|c|c|c|c}
\hline
\hline
 --			& \taumu GMSB			& \taumu nGM 				& \taumu bRPV				& \taumu mSUGRA			\\
\hline\hline
Multijet 		& \taumuGMSBQCD		& \taumuNGMQCD				& \taumuBRPVQCD			& \taumuMSugQCD			\\
$W$ + jets		& \taumuGMSBW			& \taumuNGMW				& \taumuBRPVW				& \taumuMSugW				\\ 
$Z$ + jets		& \taumuGMSBZ			& \taumuNGMZ				& \taumuBRPVZ				& \taumuMSugZ				\\ 
Top		& \taumuGMSBTop		& \taumuNGMTop			& \taumuBRPVTop			& \taumuMSugTop		
\\
Diboson			& \taumuGMSBDiBoson		& \taumuNGMDiBoson			& \taumuBRPVDiBoson			& \taumuMSugDiBoson			\\
\hline\hline
Total background 	& \taumuGMSBAll		& \taumuNGMAll	& \taumuBRPVAll	& \taumuMSugAll	\\
\hline\hline
Data       		& \taumuGMSBData		& \taumuNGMData        		& \taumuBRPVData			& \taumuMSugData        		\\
\hline
\hline
Signal MC Events &&&\\
~~GMSB 60/30		& \taumuGMSBBench 				& --					& -- 					& --					\\
~~nGM 940/210 		& -- 				& \taumuNGMBench					& -- 					& --					\\
~~bRPV 600/600 		& -- 				& --					& \taumuBRPVBench 					& --					\\
~~mSUGRA 800/400	& -- 				& --					& -- 					& \taumuMSUGRABench					\\
\hline
\hline
Obs (exp) limit &&&\\
on signal events 	& 5.3 (${4.0}^{+1.6}_{-0.2}$) & 4.6 (${5.6}^{+2.1}_{-1.5}$)		& 10.6 (${6.1}^{+2.6}_{-1.0}$)  	& 9.9 (${10.0}^{+3.6}_{-2.7}$)	\\
\hline
Obs (exp) limit on &&&\\
vis. cross section (fb) 	& 0.26 (0.20)		& 0.23 (0.28)				& 0.52 (0.30)			& 0.49 (0.49)					\\
\hline
Discovery p-value&&&\\
$p(s=0)$		& 0.22				& 0.50					& 0.04   				& 0.50				\\
\hline
\hline
\end{tabular}

\end{center}
\end{table*}

\subsection*{Interpretation}

A statistical combination of SRs is performed to produce 95\%~\CL limits on the model parameters for all signal models. For each scenario the combination of SRs from 
each channel that gives the best expected sensitivity is chosen (see \tab~\ref{tab:SRCombination}). 
In setting the limits the full likelihood function that represents the outcome of the combination is used. The combination profits 
from the fact that all channels considered in the analysis are statistically independent.
The limits are calculated using an asymptotic approximation and including all experimental uncertainties on the background and signal expectations, as well as theoretical uncertainties on the background, as nuisance parameters, neglecting any possible signal contamination in the control regions. 
Correlations between signal and background uncertainties are taken into account.

\begin{table}[ht]
 \centering
\caption{Overview of the signal regions used from each channel for the combined limit setting.}
 \label{tab:SRCombination}
 \begin{tabular}{l|cccc}
 \hline
 \hline
 Signal scenario	& \onetau SR	& \twotau SR	& \taulep SR 	\\
 \hline
 GMSB			& Tight		& GMSB		& GMSB		\\
 nGM			& --		& nGM		& nGM		\\
 bRPV			& Tight		& bRPV		& bRPV		\\
 mSUGRA			& Tight		& --		& mSUGRA	\\
 \hline
 \hline
 \end{tabular}
\end{table}

\begin{figure}[hp]
  \vspace*{-4ex}
  \begin{center}
   \includegraphics[width=0.58\textwidth,bb=0 0 567 531]{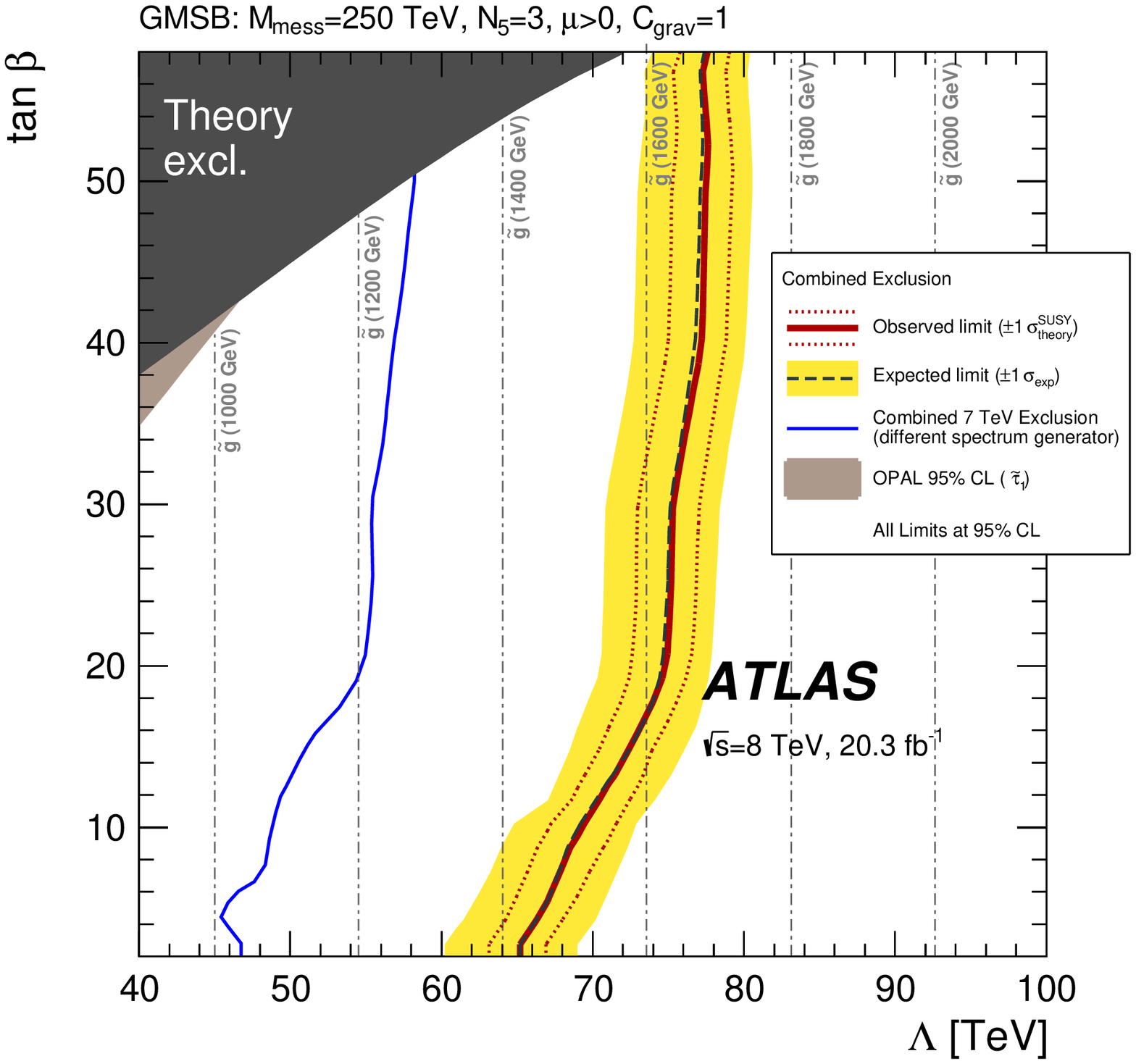}
    \vspace*{-4ex}%
  \end{center}
  \caption{Expected and observed 95\%~\CL lower limits on the
    minimal GMSB model parameters $\Lambda$ and $\tan\beta$ using a combination of all channels. 
    The result is obtained using \integLumi of $\sqrt{s} = \unit[8]{TeV}$ ATLAS data. 
    The dark grey area indicates the region which
    is theoretically excluded due to unphysical sparticle mass values.
    Additional model parameters are $\Mmess = \unit[250]{TeV}$,
    $\Nfive=3$, $\mu>0$ and $\Cgrav=1$. The
    OPAL limits on the \stau{} mass~\protect\cite{Abbiendi:2005gc}
    and the previous ATLAS~\cite{ATLAS:SusyTau5fb} limits are shown. 
    For the latter, a different mass spectrum generator was employed.}
 \label{fig:gmsb:limit:combined}
\end{figure}

The resulting observed and expected limits in the GMSB scenario for the combination of all final states considered are shown in 
\fig~\ref{fig:gmsb:limit:combined}. The yellow band around the expected exclusion limit represents the 1$\sigma$ statistical and 
systematic uncertainty on the 
expected background, as well as the experimental uncertainty on the signal. The dashed red lines around the observed limit indicate the effect of the theoretical uncertainties on the signal cross section.
The limits quoted in the following correspond to the assumption that  
the signal cross section is reduced by $1\sigma$. A lower limit on the SUSY breaking scale $\Lambda$ of \GMSBLimitAll is
determined, independent of the value of $\tan\beta$. The limit on $\Lambda$ increases to \GMSBLimitBestTanBeta for large $\tan\beta$ ($\tan\beta>20$). This corresponds to excluding gluino masses
lower than about \GMSBLimitGluino. These are the strongest available limits on GMSB-like SUSY with tau lepton signatures.

\begin{figure}[hp]
 \vspace*{1ex}%
 \begin{center}
  \includegraphics[width=0.58\textwidth,bb=0 0 567 531]{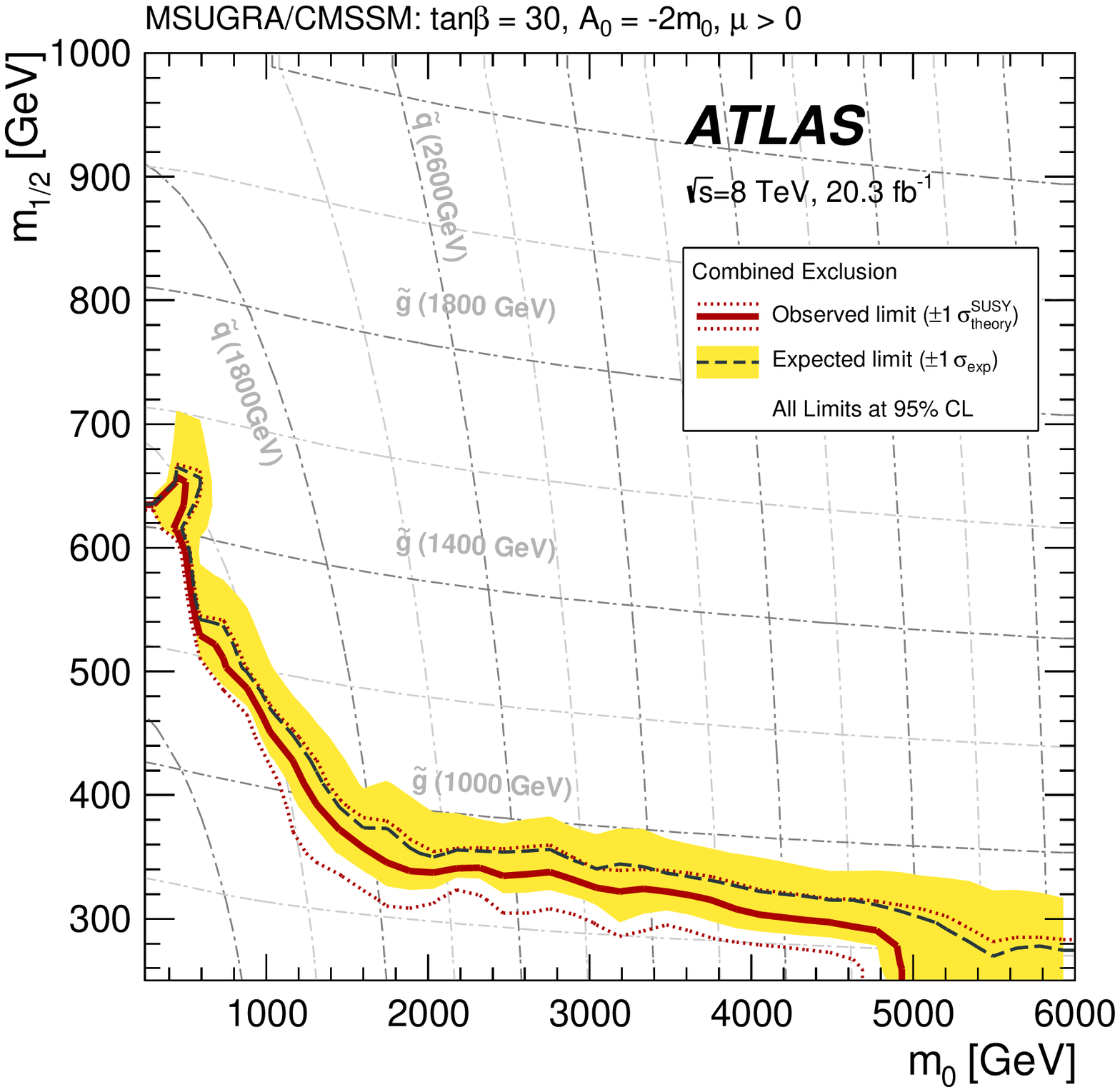}
    \vspace*{-4ex}%
 \end{center}
 \caption{Expected and observed 95\%~\CL lower limits on the
   mSUGRA/CMSSM model parameters \mz and \moh for the
   combination of the \onetau, \tauel and \taumu channels.
   Additional model parameters are $A_0 = -2\mz$,
   $\tan\beta = 30$ and $\mathrm{sign}(\mu) = +1$.
}
 \label{fig:msugra:limit}
\end{figure}

\Fig~\ref{fig:msugra:limit} shows the expected and observed exclusion limits obtained when interpreting the \onetau and \taulep analysis results in the mSUGRA/CMSSM model plane. 
Values of $m_{1/2}$ up to \mSugramohlimitsmall for low $m_0$ and \mSugramohlimitlarge for larger $m_0$ ($m_0>\GeV{2000}$) are excluded.

\begin{figure}[hp]
  \vspace*{-4ex}
  \begin{center}
   \includegraphics[width=0.59\textwidth,bb=0 0 567 531]{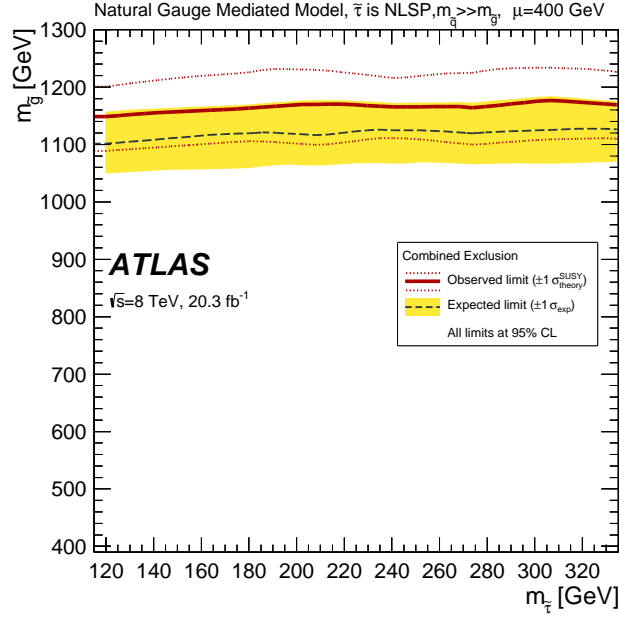}
    \vspace*{-4ex}%
  \end{center}
  \caption{Expected and observed 95\%~\CL lower limits on the
   simplified nGM model parameters $m_{\widetilde{\tau}}$ and $m_{\widetilde{g}}$
   for the combination of the \twotau, \tauel and \taumu channels.
   Additional squark and slepton mass parameters are set to \unit[2.5]{TeV},
   $M_1=M_2=\unit[2.5]{TeV}$, and all trilinear coupling terms are set to zero.
   Also, the parameter $\mu$ is fixed to $\mu=\unit[400]{GeV}$.
}
 \label{fig:ngm:limit}
\end{figure}

\Fig~\ref{fig:ngm:limit} shows the expected and observed nGM exclusion limit obtained using the dedicated SRs of the \twotau and the \taulep channels for this scenario. 
Exclusion limits on the mass of the gluino are set at \nGMLimitAll, independent of the \stau ~mass.

\begin{figure}[hp]
  \begin{center}
    \includegraphics[width=0.59\textwidth,bb=0 0 567 531]
                   {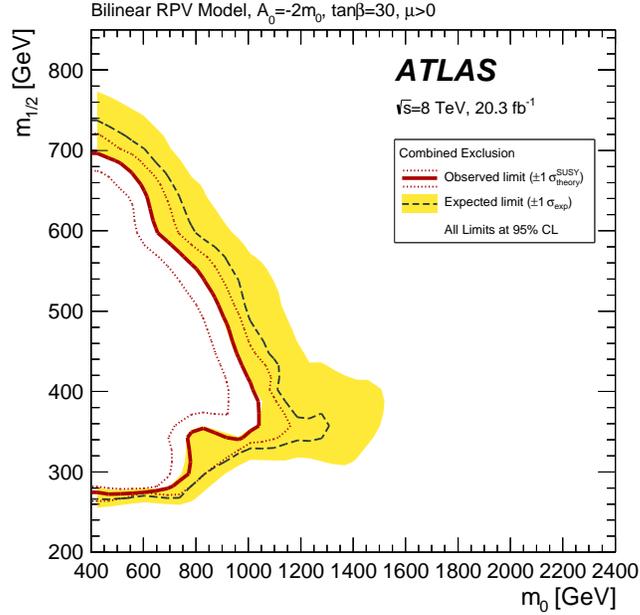}
                    \vspace*{-4ex}%
  \end{center}
  \caption{Expected and observed 95\%~\CL lower limits on the
   bRPV model parameters $m_{0}$ and $m_{1/2}$ for the combination of all channels.
   Additional model parameters are $A_0 = -2\mz$,
   $\tan\beta = 30$ and $\mathrm{sign}(\mu) = +1$.
}
 \label{fig:brpv:limit}
\end{figure}

\Fig~\ref{fig:brpv:limit} shows the expected and observed exclusion limit in the bRPV scenario for the combination of all final states considered.
Values of $m_{1/2}$ up to \bRPVmohlimit are excluded for low $m_0$, while the exclusion along 
the $m_0$ axis reaches a maximum of \bRPVmzlimit for \linebreak $m_{1/2}=\bRPVmzlimitposition$.

\FloatBarrier

\section{Conclusions}

A search for supersymmetry in final states with jets, \met{}
and one or more hadronically decaying tau leptons is performed
using \integLumi of $\sqrt{s} = \unit[8]{TeV}$ $pp$ collision
data recorded with the ATLAS detector at the LHC. No excess
over the expected Standard Model background is observed.
The results are used to set model-independent 95\%~\CL
upper limits on the number of signal events from new phenomena and corresponding
upper limits on the visible cross sections.

A limit on the SUSY breaking scale $\Lambda$ of \GMSBLimitAll is
obtained, independent of the value of $\tan\beta$, for a minimal GMSB model.
The limit on $\Lambda$ increases to $\unit[73]{TeV}$ for high $\tan\beta$ ($\tan\beta>20$).
In a natural Gauge Mediation model, a limit on the gluino mass
of $\unit[1090]{GeV}$ independent of the \stau{} mass (provided the \stau{} is the NLSP) 
is obtained from the combination of the \twotau and \taulep channels.
The results of the analysis in the \onetau and the \taulep channels are interpreted in the mSUGRA/CMSSM model and stringent limits in the (\mz,~\moh) plane are obtained.
In the bilinear R-parity-violating scenario, values of $m_{1/2}$ up to $\unit[680]{GeV}$ are excluded for low $m_0$. Moreover,
values of $m_0$ up to $\unit[920]{GeV}$ are excluded, for \mbox{$m_{1/2}=\unit[360]{GeV}$}.

%
\section*{Acknowledgements}

We thank CERN for the very successful operation of the LHC, as well as the
support staff from our institutions without whom ATLAS could not be
operated efficiently.

We acknowledge the support of ANPCy\kern-0.1em{}T, Argentina; YerPhI, Armenia; ARC,
Australia; BMWF and FWF, Austria; ANAS, Azerbaijan; SSTC, Belarus; CNPq and FAPESP,
Brazil; NSERC, NRC and CFI, Canada; CERN; CONICYT, Chile; CAS, MOST and NSFC,
Chi\-na; COLCIENCIAS, Colombia; MSMT CR, MPO CR and VSC CR, Czech Republic;
DNRF, DNSRC and Lundbeck Foundation, Denmark; EPLANET, ERC and NSRF, European Union;
IN2P3-CNRS, CEA-DSM/IR\-FU, Fra\-nce; GNSF, Georgia; BMBF, DFG, HGF, MPG and AvH
Foundation, Germany; GSRT and NSRF, Greece; ISF, MINERVA, GIF, I-CORE and Benoziyo Center,
Israel; INFN, Italy; MEXT and JSPS, Japan; CNRST, Morocco; FOM and NWO,
Netherlands; BRF and RCN, Norway; MNiSW and NCN, Poland; GRICES and FCT, Portugal; MNE/IFA, Romania; MES of Russia and ROS\-ATOM, Russian Federation; JINR; MSTD,
Serbia; MSSR, Slovakia; ARRS and MIZ\v{S}, Slovenia; DST/NRF, South Africa;
MINECO, Spain; SRC and Wallenberg Foundation, Sweden; SER, SNSF and Cantons of
Bern and Geneva, Switzerland; NSC, Taiwan; TAEK, Turkey; STFC, the Royal
Society and Leverhulme Trust, United Kingdom; DOE and NSF, United States of
America.

The crucial computing support from all WLCG partners is acknowledged
gratefully, in particular from CERN and the ATLAS Tier-1 facilities at
TRIUMF (Canada), NDGF (Denmark, Norway, Sweden), CC-IN2P3 (France),
KIT/GridKA (Germany), INFN-CNAF (Italy), NL-T1 (Netherlands), PIC (Spain),
ASGC (Taiwan), RAL (UK) and BNL (USA) and in the Tier-2 facilities
worldwide.

\clearpage

\bibliographystyle{JHEP}
\bibliography{taustrong}
%
%
\clearpage
\begin{flushleft}
{\Large The ATLAS Collaboration}

\bigskip

G.~Aad$^{\rm 84}$,
B.~Abbott$^{\rm 112}$,
J.~Abdallah$^{\rm 152}$,
S.~Abdel~Khalek$^{\rm 116}$,
O.~Abdinov$^{\rm 11}$,
R.~Aben$^{\rm 106}$,
B.~Abi$^{\rm 113}$,
M.~Abolins$^{\rm 89}$,
O.S.~AbouZeid$^{\rm 159}$,
H.~Abramowicz$^{\rm 154}$
% ...
\bigskip
\\
% ...
$^{11}$ Institute of Physics, Azerbaijan Academy of Sciences, Baku, Azerbaijan\\
% ...
$^{84}$ CPPM, Aix-Marseille Universit{\'e} and CNRS/IN2P3, Marseille, France\\
% ...
$^{89}$ Department of Physics and Astronomy, Michigan State University, East Lansing MI, United States of America\\
% ...
$^{106}$ Nikhef National Institute for Subatomic Physics and University of Amsterdam, Amsterdam, Netherlands\\
% ...
$^{112}$ Homer L. Dodge Department of Physics and Astronomy, University of Oklahoma, Norman OK, United States of America\\
$^{113}$ Department of Physics, Oklahoma State University, Stillwater OK, United States of America\\
% ...
$^{116}$ LAL, Universit{\'e} Paris-Sud and CNRS/IN2P3, Orsay, France\\
% ...
$^{152}$ Institute of Physics, Academia Sinica, Taipei, Taiwan\\
% ...
$^{154}$ Raymond and Beverly Sackler School of Physics and Astronomy, Tel Aviv University, Tel Aviv, Israel\\
% ...
$^{159}$ Department of Physics, University of Toronto, Toronto ON, Canada\\
% ...
\end{flushleft}

\end{document}